\documentclass[prx,superscriptaddress,nofootinbib,notitlepage,amsmath,amssymb,showpacs,floatfix]{revtex4-1}

\usepackage{graphicx,amssymb,amsmath,amsfonts, bm}

\usepackage{colortbl}
\usepackage{tabularx}
\usepackage{verbatim}
\usepackage{multirow}
\usepackage[T1]{fontenc} 
\usepackage[utf8]{inputenc}
\usepackage{natbib}
\usepackage{umoline}
\usepackage[usenames,svgnames]{xcolor}

\usepackage[hyperindex,breaklinks]{hyperref}
\hypersetup{
     colorlinks=true,       		
     linkcolor=Salmon,          	
     citecolor=blue,            
     filecolor=blue,      		
     urlcolor=cyan,           	
 }

\newtheorem{theorem}{Theorem}[section]

\newtheorem{definition}[theorem]{Definition}

\newtheorem{lem}[theorem]{Lemma}  
\newtheorem{lemma}[theorem]{Lemma}

\newtheorem{corol}[theorem]{Corollary}

\newcommand{\qedsymb}{\hfill{\rule{2mm}{2mm}}}  
\newenvironment{proof}[1][]{\begin{trivlist}  
  \item[\hspace{\labelsep}{\bf\noindent Proof#1:\/}]}
  {\qedsymb\end{trivlist}}

\newcommand{\Eq}[1]{Eq.~(\ref{#1})}
\newcommand{\Fig}[1]{Fig.~\ref{#1}}

\newcommand{\Lem}[1]{Lemma~\ref{#1}}

\newcommand{\Sec}[1]{Sec.~\ref{#1}}
\newcommand{\Ref}[1]{Ref.~\cite{#1}}
\newcommand{\Refs}[1]{Refs.~\cite{#1}}

\newcommand{\Thm}[1]{Theorem~\ref{#1}}

\newcommand{\cc}[1]{~\cite{#1}}

\newcommand{\av}[1]{\langle #1 \rangle}
\newcommand{\bra}[1]{\langle #1 |}
\newcommand{\ket}[1]{| #1 \rangle}

\newcommand{\Oorderof}{\mathcal{O}}
\newcommand{\orderof}[1]{\Oorderof(#1)}
\newcommand{\norm}[1]{\|#1\|}
\newcommand{\Tr}{\mathop{\rm Tr}\nolimits}
\newcommand{\gs}{\Omega}

\newcommand{\EqDef}{:=}

\newcommand{\Id}{{\mathbb{I}}}
\newcommand{\ignore}[1]{}
\newcommand{\co}{{\rm c}}

\DeclareMathOperator{\arccosh}{arccosh}

\definecolor{cl1}{rgb}{0.75,0.85,0.95}
\definecolor{cl2}{rgb}{0.88,0.93,1}
\definecolor{cl3}{rgb}{0.5,0.6,0.8}

\begin{document}


\title{Local reversibility  and
entanglement structure of many-body ground states.}


\author{Tomotaka Kuwahara}
\affiliation{Department of Physics, The University of Tokyo, 
  Komaba, Meguro, Tokyo 153-8505}
\affiliation{WPI, Advanced Institute for Materials Research, Tohoku University, Sendai 980-8577, Japan}
\author{Itai Arad}
\affiliation{Centre for Quantum Technologies, 
  National University of Singapore, 3 Science Drive 2, 
  Singapore 117543}
\author{Luigi Amico}
\affiliation{Centre for Quantum Technologies, 
  National University of Singapore, 3 Science Drive 2, 
  Singapore 117543}
\affiliation{CNR-MATIS-IMM $\&$ Dipartimento di Fisica e Astronomia, 
  Via S. Sofia 64, 95127 Catania, Italy}
\affiliation{INFN-Laboratori Nazionali del Sud, INFN, via S. Sofia 62, 95123 Catania, Italy.}
\author{Vlatko Vedral}
\affiliation{Centre for Quantum Technologies, 
  National University of Singapore, 3 Science Drive 2, 
  Singapore 117543}
\affiliation{Atomic and Laser Physics, Clarendon Laboratory, 
  University of Oxford, Parks Road, Oxford, OX13PU, 
  United Kingdom}
  


\begin{abstract}
  The low-temperature physics of quantum many-body systems is
  largely governed by the structure of their ground states.
  Minimizing the energy of local interactions, ground states often
  reflect strong properties of locality such as the area law for
  entanglement entropy and the exponential decay of correlations
  between spatially separated observables. Here, we present
  a novel characterization of  quantum states, which we
  call \emph{`local reversibility'}. It characterizes the type of
  operations that are needed to \emph{reverse} the action of a
  general disturbance on the state. We prove that {unique} ground
  states of gapped local Hamiltonian are locally reversible. This
  way, we identify new {universal} features of many-body ground
  states, which cannot be derived from the aforementioned
  properties. We use local reversibility to distinguish between
  states enjoying microscopic and macroscopic quantum phenomena.  To
  demonstrate the potential of our approach, we prove specific
  properties of ground states, which are relevant both to critical
  and non-critical theories.
\end{abstract}

\maketitle

\section{Introduction}

Gapped ground states define quantum phases of matter at zero
temperature. Even though they occupy a tiny fraction of the possible
many-body Hilbert space, these states manifest a rich and diverse
structure. Standard examples are states with local order-parameter
such as paramagnetic and ferromagnetic ground states, the superfluid
and insulator ones in bosonic and fermionic many-body systems, etc.
Other instances, such as quantum Hall and quantum spin liquids, can
arise because of more subtle orders that can be established in the
system.  A central goal of condensed matter theory is to understand
their structure and how it relates to the physics of different
phases\cc{Wen_book,ref:Wen-topo}. A natural approach to this problem
is to find the constraints that these states satisfy, which set them
apart from generic many-body states\cc{hayden2006aspects}.  Such
analysis can serve for the understanding of which type of
entanglement that ground states can indeed harbour. To this
aim, it is important to understand aspects of locality in these
states.  We ask:
\textit{`to what extent can such states be described by a
collection of local degrees of freedom, which are only loosely
correlated with each other?'}

Rigorous tools to tackle this question are scarce,
even though various properties have been known in empirical ways (see \cite{ref:Osborne-QHC12}).  
 An example is provided by  the exponential decay of
correlations, also known as exponential clustering: it has been proved
that gapped ground states on a lattice have a finite correlation
length, beyond which the correlations between spatially separated
observables decay exponentially\cc{ref:Hastings04-LSM,
ref:Hastings06-Expdec, ref:LR-Nachtergaele06}. More recently, other
quantitative tools have been devised, which characterize the ground
state's locality by looking at its entanglement
structure\cc{ref:AL-rev,ref:Ent-rev-2008}. A notable example is area
law of the entanglement entropy\cc{ref:AL-rev}, which states that
the entanglement entropy of a region with respect to the rest of the
lattice should scale like the boundary area of the region rather
than its volume. It is expected to hold for all gapped ground states
on a lattice, but has only been rigorously proved in one spatial
dimension (1D) by Hastings\cc{ref:Hastings-AL07} (see
\Refs{ref:Arad-AL12, ref:Arad-AL13, cho2014sufficient, Brandao,arad2016rigorous} for
further results). Hastings' celebrated result yields a complete
characterization of 1D gapped ground states as matrix product states
(MPS)\cc{verstraete2008matrix}, which, to a large extent,
provides a full understanding of the 1D case\cc{ref:Wen10}.

Unfortunately, in higher spatial dimensions our understanding
of the problem is still very much limited. 
Not only that a proof for the area law is lacking,
but it is also unclear how an area law would imply an efficient
representation of the ground state\cc{ref:Ge2014-AL-no-PEPS}.
Moreover, when the system has long-range interactions, or it is
hosted in a lattice with a large dimensionality (like an expander
graph\cc{hoory2006expander}), locality properties of the ground
state are even more illusive: exponential decay of correlations no
longer holds (since all particles are essentially close to each
other), and in general, area law become meaningless as surface areas
become as large as volumes. 
{For such systems, very well studied in the Hamiltonian complexity field,  spatial distance might no longer a good figure of merit for identifying entanglement~\cite{ref:KempeKitaevRegev, PhysRevB.76.035114,aharonov2013guest}.
As we will shortly show, an alternative approach is to study
entanglement and locality by analyzing the collective properties of
a subsystem with respect to the \emph{number} of local degrees of
freedom it contains rather than the \emph{distance} between them.}

In this paper, we introduce a new constraint  on a many-body gapped ground states which
complements some of the shortcomings of
the existing approaches. We call it {\it local reversibility}.
It is based on the intuition that macroscopic-scale
entanglement cannot be recovered by any local operation once it has
been broken. Therefore, states which allow this sort of local
recovery, necessarily contain a 'small amount of macroscopic superposition'. 
Here, we observe that  we use the term locality in a broader
meaning than the usual spatial locality.

We will show that such local reversibility holds for \emph{all}
unique gapped ground states of local Hamiltonians, including systems
with long-range interactions or a diverging lattice dimensionality
(for which the existing approaches to the locality properties, like
the exponential decay of correlation, do not apply). We
therefore believe that it exposes fundamental features of gapped
ground states that cannot be captured by existing properties.
To demonstrate its potential, we study specific problems in
many-body physics.  We work out rigorous bounds for the quantum
fluctuations of locally reversible states. This, in turn, implies
new constraints on the critical exponents and rigorous bounds on the
quality of the mean-field ansatz, which is often used to treat
complicated quantum many-body systems. An important outcome of our
approach is an effective way to identify quantum macroscopic 
superposition.

\section{Local reversibility} 
\begin{figure}
  \centering\includegraphics[clip, scale=0.44]{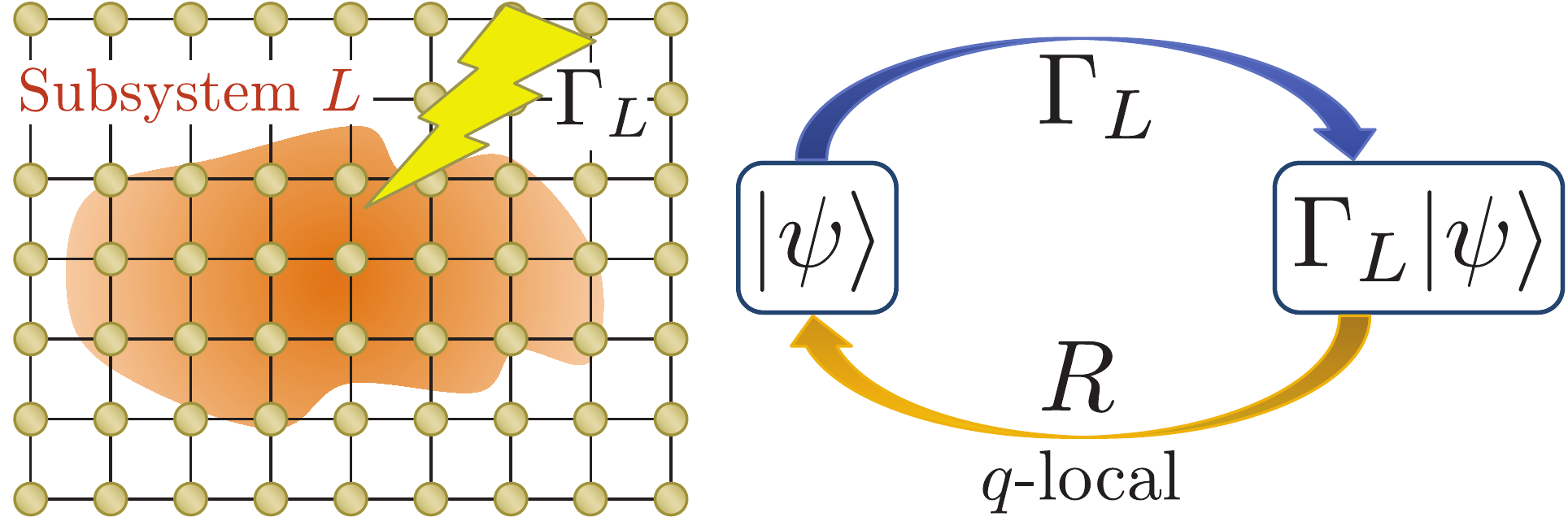}
    \caption{
    Schematic picture of the local reversibility (LR). We disturb a
    quantum state $\ket{\psi}$ by an operator $\Gamma_L$, which is
    supported in a subsystem $L$. We then try to recover the state
    $\Gamma_L \ket{\psi}$ by the use of a $q$-local operator $R$. If
    the state $\ket{\psi}$ is a product state, we can recover the
    original state by an operator $R$ with $q
    =\orderof{\sqrt{|L|}}$; then, `locally reversible state' is
    defined as the class of states which have the same property as
    product state in terms of the non-locality of the reverse
    operator.  The entanglement properties of LR states are expected
    to be highly restricted since entanglement cannot be recovered
    by local operations once it has been broken.}
    \label{fig:local_reversibility}
\end{figure}

{~}\\ \label{section:LR} 
To motivate our approach, we begin with a heuristic discussion
(Fig.~\ref{fig:local_reversibility}).  Consider a state $\ket{\psi}$ that is defined over $N$
localized spins, each with a $d$-dimensional Hilbert space, and let
$\Gamma_L$ be an operator acting on a spin subset $L$; the total
system is given by $L\cup L^\co$ with $L^\co$ the complement of $L$.
Applying $\Gamma_L$ to $\ket{\psi}$, we can potentially disrupt the
entanglement between $L$ and $L^\co$, even when $\Gamma_L\ket{\psi}$
has a constant overlap with $\ket{\psi}$. It is useful to think of
$\ket{\psi}$ as a superposition of several states
$\ket{\psi}=\ket{\psi_1}+\ket{\psi_2} + \cdots$ and of $\Gamma_L$ as
a projector that ``kills'' some (but not all) of these states.
Intuitively, if $\ket{\psi}$ contains  some ``global entanglement''  on the
scale of $|L|$ spins, we may only be able to reconstruct
$\ket{\psi}$ by acting on $\Gamma_L\ket{\psi}$ with (at least) an
operator that acts non-trivially on the same portion $L$ of the
system (i.e., it would be an $|L|$-local operator).  However, when
$\ket{\psi}$ contains mostly short-range entanglement, we might be
able to return to $\ket{\psi}$ by using an operator of a much
smaller support.  How much smaller should that support be for a
slightly entangled state? Specifically, as we shall see shortly, the
minimal size of support that is needed to reconstruct a product
state is $\orderof{\sqrt{|L|}}$. This indicates that states that
can be reversed by operators of $\orderof{\sqrt{|L|}}$ support
constitute a class of states with a small amount of entanglement. In
the following, we refer to such a class as locally reversible
states.

{We now put the discussion above on a formal ground. We first defines
the {notion} of $q$-local {operator}, which may be
often called {a} ``few-body {operator}:''
\begin{definition}[${\bm q}$-local]
  {Given an integer $q>0$, a $q$-local operator is an operator of
  the form} $O\EqDef\sum_{|X|\le q } o_X $, where {each} $o_X$
  is {an} operator supported {on} a finite
  subset of spins $X=\{i_1,i_2,\ldots, i_{|X|}\}$ {of}
  cardinality $|X|${. The $o_X$ operators are}
  \emph{not} necessarily {sitting} next to each
  other on the lattice.
\end{definition}}
We formulate the reversibility property in terms of such operators $o_X$. 
\begin{definition}[Local Reversibility]
\label{def:LR-def}
  We say that a state $\ket{\psi}$ is locally reversible (LR) if
  there exists a function $f(x)$ that decays faster than any power
  law, such that for every subset of spins $L$ and an operator
  $\Gamma_L$ defined on it, and for every integer $q>0$, there
  exists a $q$-local operator $R$ 
  such that
  \begin{align}
    \norm{R\Gamma_L \ket{\psi} -  \ket{\psi}}
    \le  \frac{\|\Gamma_L \|}{|\bra{\psi} \Gamma_L \ket{\psi}| }
      f\biggl(\frac{q}{\sqrt{|{L}|}}\biggr) \, ,
    \label{def:LR}
  \end{align}
  where $\norm{\cdots}$ is the operator norm. 
\end{definition}


{Three} remarks are in order. \textit{i)} Both the shape and the size of
$L$ are left completely general. In particular, we can take $L$ to
be the entire system ($|L|=N$).  \textit{ii)} In some cases, it will
make sense to only consider operators $R$ that respect certain
symmetries. We will later use this restricted definition of local
reversibility for states with symmetry protected topological order
(SPTO)\cc{chen2011symmetry}. { The last remark is on the status of function $f(x)$ in (\ref{def:LR}).
Despite   $f(x)$ need to be a superpolynomially decaying function in \ref{def:LR-def},   the statement (\ref{def:LR}) itself can be proved for a fixed generic $f(x)$. In this sense, the statement is non-asymptotic and valid for finite systems. In order  (\ref{def:LR}) to be effective  in  putting bounds on the state in a meaningful way, however,  $f(x)$ need to be specific and non-trivial (see our main theorem \ref{thm:main} for an example of $f(x)$); such a feature will be thoroughly exploited in the rest of the paper.}

{We claim that LR states show a specific degree of locality,
while non-LR states correspond to states with non-local
features due to global entanglement.  This assertion
can be  explained  by the following two lemmas characterizing the entanglement structure of LR states.}

{The first lemma refers to the so-called macroscopicity of the states. Namely, we will demonstrate how non-LR states correspond to states with  macroscopic superposition.}

{Let's consider states of the type
$\ket{\psi}=\alpha\ket{\psi_a} + \beta\ket{\psi_b}$ and discuss the possibility that 
$\ket{\psi_a}$ and $\ket{\psi_b}$ are macroscopically distinct (meaning that a  collection of local operators  exists to  $\ket{\psi_a} \leftrightarrow \ket{\psi_b}$).}
%
Then:
\begin{lemma}
\label{lem:macro}
{Let $\ket{\psi}$ be a state which satisfies~\eqref{def:LR} for a fixed function $f(x)$. 
Then, for any decomposition  $\ket{\psi}= P_L\ket{\psi} + (\Id-P_L)\ket{\psi}\EqDef \alpha\ket{\psi_a} + \beta\ket{\psi_b}$ with $P_L^2=P_{L}$, 
we have
   \begin{align*}
  O\ket{\psi_a} =  \ket{\psi_b} + \ket{\delta},
  \end{align*}
 where  $O$ is a $q$-local operator and   ${\|\delta\|^{2}}:=\langle\delta \ket{\delta}  \le|\alpha^2\beta|^{-1} f(q/\sqrt{|L|})$. 
 When $\ket{\psi}$ is a LR state, $f(x)$ decays superpolynomially and only a  difference of $\orderof{\sqrt{|L|}}$ exists between  the two states  $\ket{\psi_a}$ and $\ket{\psi_b}$. } 
\end{lemma}
The proof is provided in the Appendix \ref{proofmacro}.

{By contraposition of  the lemma}, any quantum state such that we can find a bipartition $P_L$ for which two states 
$\ket{\psi_a}$ and $\ket{\psi_b}$ are macroscopically distinct over a $\sqrt{|L|}$ spatial scale, is non-LR; for example,  
the GHZ state over $n$
particles, $\ket{\psi}=\frac{1}{\sqrt{2}}(\ket{0\cdots 0}_n +
\ket{1\cdots 1}_n)$ is not LR since $\ket{0\cdots
0}_n,\ket{1\cdots 1}_n$ are clearly macroscopically distinct over
the scale of $n-1$, and we may write $\ket{\psi}=P_L\ket{\psi}
+ (\Id-P_L)\ket{\psi}$ with $P_L\EqDef \ket{0\cdots
0}\bra{0\cdots 0}_n$.
  As we show later, this simple lemma also shows
that degenerate topologically ordered states  are not LR.





The second lemma  shows that fluctuations in an LR state are
strongly suppressed. Indeed, consider an LR state $\ket{\psi}$
together with a subset of spins $L$, and let $A_L$ be an additive
operator of the form $A_L\EqDef \sum_{i\in L}a_i$. Here, each $a_i$
is an Hermitian operator with $\norm{a_i}\le 1$, which acts only on
the $i$th spin. 
Since the $a_i$ operators are commuting with each
other, they can be viewed as classical random variables whose joint
probability distribution is given by the underlying state
$\ket{\psi}$. The following lemma shows that their sum resembles a
sum of \emph{independent} random variables: its probability
distribution is strongly concentrated around its mean with a
width of $\orderof{\sqrt{|L|}}$.
\begin{lem}
\label{lem:fluc} 
  Let $\Pi^{A}_{\le x}$ and $\Pi^{A}_{> x}$ be the projectors onto
  the eigenspaces of $A_L$ with eigenvalues $\le x$ and $>x$
  respectively, and let $m$ be the median of $A_L$ with respect to
  $\ket{\psi}$ {satisfying \eqref{def:LR}}  in the sense that $\bra{\psi}\Pi^{A}_{\le
  m}\ket{\psi} \ge 1/2$ and $\bra{\psi}\Pi^{A}_{\ge m}\ket{\psi} \ge
  1/2$. Then, for any positive $h$  the following inequality holds: 
  \begin{align}
  \label{eq:LR-fluc}
    \norm{ \Pi^{A}_{\ge m+h}\ket{\psi}}
      \le 2f\biggl( \frac{\lceil h/2 \rceil-1}{\sqrt{|L|}}\biggr) \,.
  \end{align}
  with a  fixed function $f(x)$. 
  An equivalent statement is  valid for $\norm{ \Pi^{A}_{\le m-h}\ket{\psi}}$.
\end{lem}
The proof is  given by choosing $P_L= \Pi^{A}_{\le m}$ in Lemma~\ref{lem:macro}. After a short algebra, we get
$
 \norm{ \Pi^{A}_{\ge m+h}\ket{\psi}} \le  |\beta| \cdot  \| \Pi^{A}_{\ge m+h} O\Pi^{A}_{\le m}\| + 2 f(q/\sqrt{|L|})
$
with $O$ $q$-local, where we use the facts $|\alpha|^2 = \bra{\psi}\Pi^{A}_{\le m}\ket{\psi} \ge 1/2$ and  $ \|\Pi^{A}_{\ge m+h}\ket{\psi_b}\|=  \norm{ \Pi^{A}_{\ge m+h}\ket{\psi}}/\beta$.
To finish the proof we will show that $\norm{\Pi^{A}_{\ge m+h}O \Pi^{A}_{\le m}}=0$ for $q<h/2$. This follows from
  the fact that $A_L$ is a sum of (commuting) 1-local operators of
  norm 1, and therefore every $q$-local operator can take an
  eigenvector $\ket{a}$ of $A_L$ with eigenvalue $a$ to a
  superposition of eigenvectors $\sum c_{a'}\ket{a'}$ with
  $|a'-a|\le 2q$. Thus, choosing $q=\lceil h/2 \rceil-1$ proves the lemma. 

{An immediate consequence of \Lem{lem:fluc}  with the assumption that $f(x)$} is a super polynomially decaying function, is that the fluctuations
of every additive operator $A_L$, which is defined on the entire
systems ($|L|=N$) must satisfy
\begin{align}
\label{eq:fluc}
  \av{(\Delta A_L)^2} \EqDef \bra{\psi}A_L^2\ket{\psi} -
  \bra{\psi}A_L\ket{\psi}^2 \le \orderof{ N }\, .
\end{align}
We point out  that the well-known notion of macroscopicity measured by the \emph{Fisher information}\cc{ref:Shimizu05-macroE,frowis2012} is implied by the Lemma~\ref{lem:macro} and Lemma~\ref{lem:fluc}.
This feature emerges clearly from the following reasoning. 
The Fisher information of a pure state $\ket{\psi}$ with respect to an
operator $A$ is given by $\mathcal{F}(\psi,A)=4\av{(\Delta
A)^2}$\cc{frowis2012}. In \Ref{frowis2012} the
authors suggest to define the `effective macroscopic size' of a
state as
$N_{\text{eff}}(\psi)\EqDef \max_A\mathcal{F}(\psi,A)/(4N)$, where
the maximization is over all extensive operators $A\EqDef \sum_i
a_i$ as in \eqref{eq:fluc}.  States showing maximal quantum
macroscopicity, such as the GHZ state, have
$N_{\text{eff}}=\orderof{N}$, whereas states with no quantum
macroscopicity have $N_{\text{eff}}=\orderof{1}$.
Inequality~\eqref{eq:fluc} therefore implies that LR states have
$N_{\text{eff}}=\orderof{1}$.
Equivalently, states with $N_{\text{eff}}=\orderof{N^p}$ for $p>0$
are necessarily non-LR.

On the other hand, the converse is \textit{not true}: there are states with 
$N_{\text{eff}}=\orderof{1}$ that are also non-LR. 
For instance, as we shall see, degenerate topologically ordered states turn out non-LR, but  still satisfy the inequality~\eqref{eq:LR-fluc}, namely $N_{\text{eff}}=\orderof{1}$.
Thereby, LR provides us a more stringent  characterization of the macroscopic superposition encoded in a many-body state.

%
%
%
%
%
%

%
%
%
%
%
%
%
%
%
%
%
%
%


\section{Reversibility of ground states} {~}\\ We now
introduce our main tool for identifying LR states.  The following
theorem states that unique gapped ground states of local
Hamiltonians are LR. It holds for a very wide class of quantum
systems that are described by $k$-local
Hamiltonians of the form
\begin{align}
\label{def:H}
  H=\sum_{|X|\le k} h_X\quad {\rm with}
    \sum_{X:X\ni {i}} \norm{h_X}  \le  g \quad \forall i \,,
\end{align}  
where $g$ is a constant of $\orderof{1}$. Note that $k$ is
not necessarily equal to $q$ from the definition of the operator $R$
above. {Also note that we  implicitly assume that the
spins sit on a lattice, but we make no direct use of the lattice
structure or its dimensionality. Instead, we use the second
condition in \eqref{def:H}, meaning that the total strength of all
interactions in which the $i$th spin participates is bounded by a
constant of $\orderof{1}$.  This definition of $H$ captures a very
wide class of quantum systems: with short-range interactions such as
the the $XY$ model, the Heisenberg model~\cite{ref:Heisen} and the
AKLT model~\cite{ref:AKLT}, as well as models with long-range
interactions such as the Lipkin-Meshcov-Glick model~\cite{LMG1}.
Typically, we have $k=2$ (i.e., two-body interaction), but several
exceptions exist {such as the 1D cluster-Ising
model~\cite{ref:Cluster_ising} ($k=3$), the toric code
model on a square lattice~\cite{ref:Kitaev03-toric} ($k=4$) 
and the string-net model on a honeycomb lattice~\cite{ref:Levin-Wen} ($k=12$). }
We denote the ground state of
$H$ by $\ket{\gs}$, and fix its energy to be $E_0=0$. The rest of
the energies are denoted by $0=E_0< E_1\le E_2 \le \cdots$. Finally,
we let $\delta E\EqDef E_1-E_0$ be the spectral gap just above the
ground state.  With this notation at hand, our main theorem is given
as follows.


\begin{theorem}
\label{thm:main} 
  With the above notations, for every spin subset $L$ and every
  operator $\Gamma_L$ defined on it, and for any positive integer
  $q$, there exists a $q$-local operator $R$ that satisfies
  \begin{align}
    \norm{R\Gamma_L\ket{\gs} -  \ket{\gs}}
      \le  \frac{6\norm{\Gamma_L}}{|\bra{\gs}\Gamma_L\ket{\gs}|}
        e^{-2n_0 / \xi },  \label{eq:LR1}
  \end{align}
  where $n_0\EqDef \lfloor q/k \rfloor$ and
  \begin{align}
    \xi\EqDef \sqrt{1+\frac{2E_c}{\delta E }} ,\quad  
    E_c={g|L|} + 8gk n_0 .
    \label{def:Ec}
  \end{align}  
\end{theorem}
Inequality~\eqref{eq:LR1}, together with the definitions of $n_0$
and $\xi$, implies that $\norm{R \Gamma_L \ket{\gs} - \ket{\gs}} \le
\frac{6\norm{\Gamma_L}}{|\bra{\gs}\Gamma_L\ket{\gs}|}
e^{-\orderof{q\sqrt{\delta E/|L|}}}$, and therefore $\ket{\gs}$ is
LR when $\delta E=\orderof{1}$.  Hence the existence of a spectral
gap places strong restrictions on structure of the ground states for
very wide class of Hamiltonians. 
{We note that the theorem requires no assumption on a spectral gap or the
size of $|L|$ and $N$; 
hence, the theorem is {\it not asymptotic} and applicable for
\textit{arbitrary} ground states in finite systems.}

\begin{figure}
  \centering \includegraphics[clip, scale=0.55]{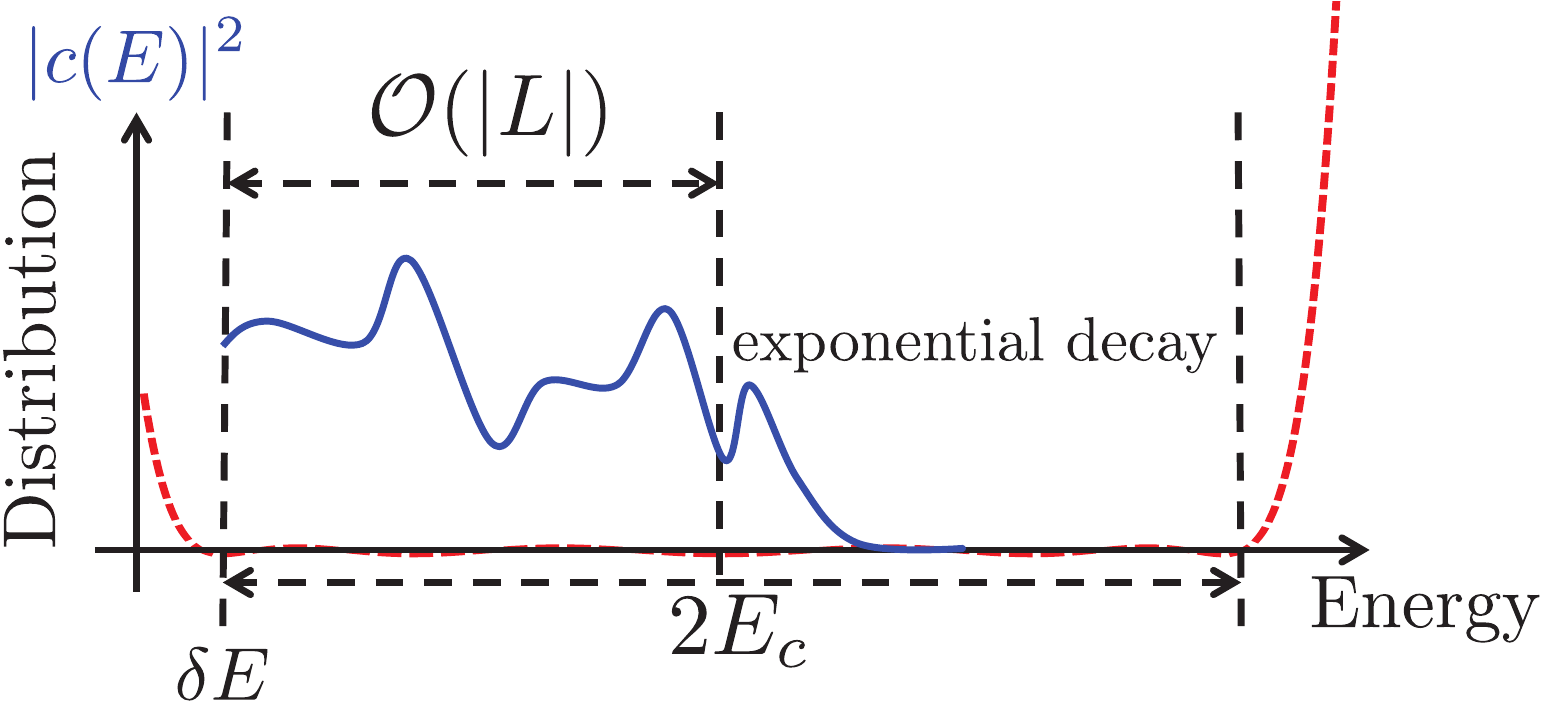}
  \caption{\footnotesize Schematic picture of the proof. After
    applying the operator $\Gamma_L$ to the ground state
    $\ket{\Omega}$, the energies at most of order $\orderof{|L|}$
    are excited (blue curve).
    We then filter out the excited states by an approximate boxcar
    function in the range $[\delta E, 2E_c+\delta E ]$ (red curve).
    Although the function rapidly increases for $x\ge 2E_c+ \delta
    E$, this can be cancelled by the exponential decay of the energy
    excitation.}
\label{fig:schematic_picutre_proof_idea_LR}
\end{figure}

The full proof of \Thm{thm:main} is given in Appendix~\ref{sec:main-proof}. 
Here we summarize its main ideas. Using recent results from
\Ref{ref:Arad14-Edist}, we conclude that after applying the operator
$\Gamma_L$ to the ground state $\ket{\gs}$, we get a state which
consists mainly of excitations with energies of at most
$\orderof{|L|}$. Beyond that scale, the weight of the excitations
decays exponentially. This is shown schematically by the blue curve
in \Fig{fig:schematic_picutre_proof_idea_LR}. Then following ideas
from a recent new proof of the 1D area law\cc{ref:Arad-AL13}, we
construct the operator $R$ by approximating the ground-state
projector using a polynomial of $H$. This polynomial is essentially
a scaled version of the Chebyshev polynomial (red curve in
\Fig{fig:schematic_picutre_proof_idea_LR}), chosen such that it
approximately behaves as a boxcar function in the range $[\delta E,
2E_c+\delta E]$, thereby suppressing the majority of excitations in
$\Gamma_L\ket{\gs}$. Crucially, even though it rapidly increases for
$x\ge 2E_c+ \delta E$, this blowup is cancelled by the exponential
decay of the high-energy excitation.


%
%

\section{Examples of locally vs. non-locally reversible
states} \label{examples}
{~}\\ Let us now apply {Lemmas~\ref{lem:macro}, \ref{lem:fluc} and \Thm{thm:main}} to
several exemplary states emerging in different contexts. The list of
states is summarized in Table~1.  In particular, we will demonstrate
how local reversibility implies the absence of macroscopic
superposition. We begin with LR states.

\noindent\textbf{1. Product states.} A product state
  $\ket{\psi}=\ket{\psi_1}\otimes\ket{\psi_2}\otimes
  \cdots\otimes\ket{\psi_N}$ is LR because it is the unique ground
  state of the local Hamiltonian $H=\sum_{i=1}^N
  (\Id-\ket{\psi_i}\bra{\psi_i}\otimes \Id_{\rm rest})$. As $H$ is
  made of commuting projectors, its spectral gap is necessarily
  $\delta E=1$. 

\noindent\textbf{2. Graph states with bounded degree.} These states
  are defined on a graph in which each node has at most
  $\orderof{1}$ neighboring nodes\cc{raussendorf2001one,
  briegel2001persistent} .  The graph state is a non-degenerate
  gapped ground states of a Hamiltonian which is the summation of
  the following commuting stabilizers\cc{hein2004multiparty}
  $\{g_i\}_{i=1}^N$: $ {g_i = \sigma_i^x\otimes
  (\sigma_{j_1}^z\sigma_{j_2}^z\cdots \sigma_{j_{k_i}}^z),} $ where
  $[g_i,g_{i'}]=0$ for $\forall i,i'$,
  $\{\sigma^{x},\sigma^{y},\sigma^{z}\}$ are the Pauli matrices and
  $\{j_1,j_2,\ldots,j_{k_i}\}$ are nodes which connect to the
  node~$i$.  By assumption, $k_i=\orderof{1}$, and hence the
  Hamiltonian is $\orderof{1}$-local. By the commutativity of its
  terms, we conclude that it has a spectral gap $\delta
  E=\orderof{1}$, and so by \Thm{thm:main} such graph states are LR.

\noindent\textbf{3. Short-range entanglement (SRE) states.} The
  third example are states that can be obtained by a constant-depth
  quantum circuit acting on a product state. In the literature they
  are often dubbed as ``trivial
  states''\cc{ref:Hastings10-nonZeroT, ref:Hastings14-NLTS}, or
  ``short-range-entanglement (SRE) states''\cc{ref:Wen-topo}. A
  constant-depth quantum circuit is a unitary operator that can be
  written as a product of $k=\orderof{1}$ unitary operators
  $U=U_1\cdots U_k$ where each unitary $U_i$ is given as a product
  of unitary operators $U_i = U_{i,1}\cdot U_{i,2}\cdots U_{i,n_i}$
  with non-overlapping support of $\orderof{1}$. To see why these
  are LR states, we write $\ket{\psi} = U\ket{\phi}$, where $U$ is
  the constant-depth circuit, and
  $\ket{\phi}=\ket{\phi_1}\otimes\ket{\phi_2}\otimes \cdots$ is a
  product state. Then it is easy to see that for any operator $O$
  with a support of $\orderof{1}$, $U O U^{-1}$ has also an
  $\orderof{1}$ support, and therefore if $H$ is a local Hamiltonian
  for which $\ket{\phi}$ is the unique ground state (see the first
  example), then $H'=UHU^{-1}$ is also a local Hamiltonian.
  Furthermore, $H'$ has the same spectrum as $H$, and so it is
  gapped with the unique ground state, which is exactly
  $\ket{\psi}$. By \Thm{thm:main} this state is LR. 
  
  We note that not all LR states are also SRE states, or,
  equivalently, long-range entanglement (LRE) does not necessarily
  imply non-LR. For example, Kitaev's toric
  code\cc{ref:Kitaev03-toric} on a sphere is a commuting local
  Hamiltonian and has a non-degenerate ground state with an 
  $\orderof{1}$ gap, and therefore by \Thm{thm:main} it is LR.
  Nevertheless, it cannot be generated by a constant depth circuit
  working on a product state, and is therefore not an SRE
  state\cc{ref:Toric-on-a-sphere}. This point is also explained in Appendix~\ref{sec:toric}.

\renewcommand{\arraystretch}{1.2}
\renewcommand{\arrayrulewidth}{1.2pt}

\begin{table*}[tt]
\label{tab:LR-vs-nonLR}
  \begin{center}
    \begin{tabular}{cc}
\multicolumn{2}{c}{\cellcolor{cl1}\textbf{Table~1: Locally Vs non-locally reversible states } \rule[-8pt]{0pt}{22pt}  }  \\
\cellcolor{cl2}LR&\cellcolor{cl2}Non-LR\\
\arrayrulecolor{cl3}\hline
\cellcolor{cl2}Product state&\cellcolor{cl2}GHZ state\\ 
\cellcolor{cl2}Bounded-degree graph states&\cellcolor{cl2}States with large fluctuation\\
\cellcolor{cl2}Short-range entangled state&\cellcolor{cl2}
      Degenerate, topologically ordered ground states \\
           \cellcolor{cl2}&\cellcolor{cl2} Degenerate, SPTO states (Symmetry-restricted non-LR)  
    \end{tabular}
   %
  \end{center}
   
\end{table*}

We now turn to non-LR states. We will use
Lemmas~\ref{lem:macro} and \ref{lem:fluc} to identify such states.

{
\noindent{\textbf{5. ``Schr\"odinger Cat'' like states}.   
  States like the GHZ are not LR by \Lem{lem:macro}.
}}

{
\noindent\textbf{6. States with Fisher information of
  $\orderof{N^p}$ with $p>1$.}   
  {As we already mentioned,} {this} 
  result comes directly from \Lem{lem:fluc}.  {Also here, a}
  quintessential example of this class is the GHZ
  state\cc{frowis2012}, which has the scaling with $p=2$.
  Moreover, the ground states at critical point are typically non-LR since they have $p=1+(2-\eta-z)/D$ (see Appendix~\ref{sec:critical}),
  where $z$ is the dynamical critical exponent, $\eta$ is the anomalous critical exponent, and $D$ is the dimension of the system.
  For example, the critical point of the 1D transverse Ising model has $z=1$ and $\eta=1/4$, which yields $p=7/4$.}

{
\noindent\textbf{7. States with degenerate topological order.}
  {While the local fluctuations in \Lem{lem:fluc} (as well
  as the} Fisher information{)} cannot detect a locally hidden
  order such as the topological order, {we can use \Lem{lem:macro} to see
  that states with a degenerate topological order are not LR}.
  We demonstrate this point {using} Kitaev's toric code
  model on a torus\cc{ref:Kitaev03-toric} with
  $\sqrt{n}\times\sqrt{n}$ sites.  {The} idea
  is that by taking $L$ to be a non-trivial loop in the
  torus of size $\sqrt{n}$, there exists an operator $T_L$
  that takes one ground state $\ket{\gs_1}$ to another ground state
  $\ket{\gs_2}$, i.e., $\ket{\gs_2}=T_L\ket{\gs_1}$. The properties
  of the topological order guarantee that for any observable $O$
  that is supported on less than $\sqrt{n}$ sites (the size of a
  Wilson loop), $\bra{\gs_1}O\ket{\gs_1} = \bra{\gs_2}O\ket{\gs_2}$
  and $\bra{\gs_1}O\ket{\gs_2} = 0$. Therefore, we may invoke
  \Lem{lem:macro} with $P_L\EqDef (\Id-T_{L})/2$, such that
  $\ket{\gs_1} = P_L\ket{\gs_1} +
  (\Id-P_L)\ket{\gs_1}\EqDef \alpha\ket{\gs_+} +
  \beta\ket{\gs_-}$, where $\ket{\gs_\pm} =
  \frac{1}{\sqrt{2}}(\ket{\gs_1}\pm\ket{\gs_2})$. It is easy to
  verify from the above properties that $\ket{\gs_+}, \ket{\gs_-}$
  are macroscopically distinct over a scale of $\orderof{n^{1/4}}$
  (they are in fact distinct over a scale of $\sqrt{n}$, i.e., the size
  of $L$), and therefore by \Lem{lem:macro}, these degenerate ground
  states are not LR.}  
  
  {We remark that non-degenerate topological order (e.g., in the toric code on the surface)  results LR (from Theorem~\ref{thm:main}). In this context, we observe that, despite the  topological entropy is non vanishing for  both degenerate and non degenerate topologically ordered ground states,  the two cases are {clearly distinct in terms of the irreducible multiparty correlation} (the issue has been recently addressed in Refs~\cite{irred:Zhou,Kato:cor,liu2014irreducible};  see also Appendix~\ref{sec:toric}): Being our approach able to detect a  `fine structure' in the nature of the multipartite correlations,  LR tells degenerate topological order apart  from non-degenerate topological order.}

\noindent\textbf{8. States with a degenerate symmetry protected 
    topological order.}
  The same arguments showing that degenerate topologically ordered
  states are not LR can be applied to the case of degenerate
  symmetry protected topological order. Such states show topological
  order only to a restricted set of operators defining a certain
  symmetry $G$\cc{chen2011symmetry}. They cannot be adiabatically
  connected to a product state using only operators from $G$, and in
  that restricted sense they are not SRE (see the following
  for the definition). An important example of states with SPTO can
  be obtained from graph state's Hamiltonian on an open lattice,
  where one removes the boundary stabilizers. This removal
  introduces degeneracy to the groundspace. Much like the case of
  Kitaev's toric code, we can also show here that the resulting
  ground states are non-LR as long as we restrict the operator $R$
  to satisfy the symmetry of the graph Hamiltonian without the
  boundary stabilizers. We refer to these states as \emph{symmetry
  restricted non-LR} states. We present an example of such states
  for 1D case\cc{son2012topological} in the Appendix \ref{srLR}.



\section{Fluctuations in locally reversible states} {~}\\
\Thm{thm:main}, together with \Lem{lem:fluc}
provides a remarkable insight into the structure of unique ground
states\footnote{{We notice that because \Thm{thm:main} is not asymptotic, the results in this section can be applied to arbitrary system size.}}.
{For any such  ground state
$\ket{\gs}$, and for any
additive operator $A_L=\sum_{i\in L}a_i$ defined on a spin subset
$L$,
$  \norm{\Pi^{A}_{\ge m+h}\ket{\gs}} 
    \le e^{-c_1 h \sqrt{\delta E/ |L|}} \, ,
$ 
with $c_1$ a constant of $\orderof{1}$ (with $m$ as defined in \Lem{lem:fluc}).}
This  implies
that $|\av{A_L}-m|=\orderof{|L|/\delta E}$, where
$\av{A_L}=\bra{\gs}A_L\ket{\gs}$ is the expectation of $A_L$ in the
ground state, and therefore 
\begin{align}
\label{eq:gs-localization}
  \norm{\Pi^{A}_{\ge \av{A_L}+h}\ket{\gs}} 
    \le e^{-c_2 h \sqrt{\delta E/ |L|}} \,,
\end{align}
{$c_2$ being a constant depending  on the Hamiltonian's parameters $k$ and $g$.}
Taking $A_L$ to be an order parameter (i.e., the
magnetization in $L$), we arrive at the conclusion that \emph{the
deviations of any order parameter from its expectation are exponentially
suppressed in unique gapped ground states.} It is interesting to
contrast this inequality with the corresponding statistics of a
product state. In such a case, $A_L$ can be viewed as a sum of
\emph{independent} random variables, and by the Hoeffding's inequality~\cc{ref:Hoeffding-ineq},
$\norm{\Pi^{A}_{\ge \av{A_L}+h}\ket{\psi}} \le
e^{-\orderof{h^2/|L|}}$. In this sense, unique gapped ground states
enjoy a weaker, yet still non-trivial, notion of local independence.

It is also worth noting that this independence cannot be (at least
directly) deduced from the exponential decay of correlation of
gapped ground states\cc{ref:Hastings06-Expdec, ref:LR-Nachtergaele06}, since it can be
applied to sets of observables that may sit very close to each other
on the lattice.  Moreover, we can apply it to systems with
long-range interactions, such as the Lipkin-Meshcov-Glick
model\cc{LMG1} and systems defined on the expander
graphs\cc{hoory2006expander}, in which the maximal distance
between any two spins is $\orderof{1}$ and $\orderof{\log N}$,
respectively.  We remark that
 inequality~\eqref{eq:gs-localization} can be extended to generic few-body operators~\cite{JSTAT_extend}: $A=\sum_{|X|\le q}  a_{X}$ with $q=\orderof{1}$; finally we can derive a similar bound for low-lying energy
states, i.e., not necessarily the exact ground state (T.K., I.A.,
L.A. and V. V., manuscript in preparation).

A simple consequence of 
inequality~\eqref{eq:gs-localization} is a trade-off relationship
between the spectral gap and the fluctuation 
{$\Delta A_L \EqDef (\bra{\gs}A_L^2\ket{\gs} -
\bra{\gs}A_L\ket{\gs}^2)^{1/2}$ of $A$} in the
ground state:
\begin{align}
  \delta E \cdot (\Delta A_L)^2 \le {\rm const} \cdot  |L|,.
    \label{eq:fluctuations}
\end{align}
This has two interesting implications:

\noindent\textbf{1. Bounds on the critical exponents.}
{As noted above,  Theorem \ref{thm:main}  does not assume the spectral gap of $\orderof{1}$ and  therefore can be applied to  \textit{arbitrary} ground states.
Below, we apply
{it} to quantum critical points to obtain a
general inequality for critical exponents.} 

Let us
consider the critical regime, $\delta E\to 0$. Define
$A_L=\sum_{i=1}^N a_i$ with $L$ a total system and $\{a_i\}_{i=1}^N$
order parameters (e.g. magnetization).  We then introduce the
critical exponents $z$, $\eta$, $\gamma$ and $\nu$ as in
\Refs{ref:Conti94-Quantum-scaling}; $z$ is the dynamical critical
exponent, $\eta$ is the anomalous critical exponent, $\gamma$ is the
susceptibility critical exponent and $\nu$ is the correlation length
exponent. By applying the finite-scaling
ansatz\cc{ref:Conti94-Quantum-scaling} 
to~\eqref{eq:fluctuations}, we can obtain
\begin{align}
\label{eq:critical-exp}
  z \ge 1-\frac{\eta}{2}=\frac{\gamma}{2\nu},
\end{align}
where the second equality comes from the Fisher equality
$2-\eta=\gamma/\nu$. 
We remark that (\ref{eq:critical-exp}) holds for very general
settings both for homogeneous and disordered critical systems
(see~\cite{vojta2005percolation} for a non-trivial example where our
inequality can be applied).  Incidentally, we note
that~\eqref{eq:fluctuations} gives non-trivial bounds for the
critical Lipkin-Meshcov-Glick model, a system with long-range
interactions\cc{LMG1,dusuel2004finite}.  The details of this calculation are given in
Appendix~\ref{sec:critical}

\noindent\textbf{2. Validity of mean-field approximations.} 
{Under the assumption of inequality~\eqref{eq:fluctuations}
for ground states, we can estimate the validity of the mean-field
approximation.} Just as the first implication, the full details are
given in Appendix~\ref{sec:MF}. The idea is that since
the operators $A_L$ in \eqref{eq:fluctuations} are arbitrary (as
long as they are additive on $L$), we can use them to probe the
two-spin reduced density matrix $\rho_{ij}$ and its relation with
its mean-field approximation $\rho_i\otimes\rho_j$. Specifically, it
can be shown that for every spin subset $L$ and an arbitrary spin
$i$ \emph{outside} of it,
\begin{align}
\label{eq:meanF}
  \sum_{j\in L}\norm{\rho_{ij}-\rho_i\otimes\rho_j}
    \le {\rm const} \cdot  \sqrt{|L|/\delta E} \,.
\end{align}
This implies that \emph{on average}, for each spin $j\in L$,
$\norm{\rho_{ij}-\rho_i\otimes\rho_j} \le \orderof{1/\sqrt{|L|\delta
E}}$. If our system is defined by a nearest-neighbor two-body 
Hamiltonian on a regular grid with coordination number $Z$ (the
number of neighbors of each spin), then taking $L$ to be the set of
neighbors ($|L|=Z$){, one} immediately obtains a bound on
the quality of the mean-field approximation for the energy density
for $\forall i$:
\begin{align*}
  \biggl |\frac{1}{Z}\sum_{<i,j>} \av{h_{ij}}_{\rm MF} 
    - \frac{1}{Z}\sum_{<i,j>} \av{h_{ij}}_{\rm exact}\bigg| 
      \le {\rm const} \cdot \frac{1}{\sqrt{Z\delta E} } \,,
\end{align*}
where the sum is taken over the spins adjacent to $i$. We therefore
obtain a quantitative bound on how the error of the mean-field
approximation decreases as the lattice dimension (on which the
coordination number depends) goes to infinity.  This result is
consistent with the folklore knowledge in condensed-matter physics
that the mean-field becomes exact in infinite dimension. Recently,
similar results have been obtained in different manners by
Brand{\~a}o~\textit{et al.}\cc{ref:Brandao13-qpcp} and
Osterloh~\textit{et al.}\cc{ref:Osterloh14-MF} {In
Ref.\cc{ref:Brandao13-qpcp}, the setup is more general {(i.e., the system is not assumed to be gapped)} but
the {error} estimation is  weaker than
ours, {scaling} as $\orderof{Z^{-1/3}}$: { In}
Ref.\cc{ref:Osterloh14-MF}, the {error} estimation is as good
as ours, $\orderof{Z^{-1/2}}$, but {under the additional} 
assumptions of {having a} regular, isotropic, and bipartite
lattice of $\frac{1}{2}$-spins.}


\section{Summary and open questions} 

In this work, we introduced a new notion of locality in quantum
states, the local reversibility, which is defined in terms of the
type of local operations that are needed to reverse the action of
perturbations to the state. 
%
%

{We proved that all unique ground states of gapped local Hamiltonians
are locally reversible (\Thm{thm:main}), and, on the other hand, we
showed how local reversibility implies a suppression of quantum
fluctuations (\Lem{lem:fluc}). Together, these two results provide
new insights into the structure of unique ground states of gapped
local-Hamiltonians: \textit{i)} a low Fisher information, which is an
indication for the lack of quantum macroscopicity in these states;
\textit{ii)} a novel inequality for the critical exponents in these
systems; \textit{iii)} a quantitative analysis of the mean-field
approximation; and finally, \textit{iv)} since an adiabatic (local
unitary) evolution of product states is locally reversible, our
result clearly implies that all the gapped quantum phases of matter,
disordered or with local order parameter (Landau symmetry breaking
quantum phases), are reversible. In contrast, degenerate topological
phases or the symmetry protected topological phases, are not
reversible. We note that LR can detect the difference between degenerate and non degenerate topological order. Indeed, it was  discovered that, although both  with non vanishing  topological entropy they have very different irreducible multipartite correlation (see paragraph 8 of Sect\ref{examples} and the Appendix \ref{sec:toric}). In this context, we observe that LR can be further restricted (with a similar logic we pursued in this article to deal with symmetry protected topological phases) to improve and refine the characterisation of the ground state. Such a strategy might lead to catch properties of the state originating from the geometry of its ambient space}.

Our work provides an instrumental view for several research 
directions.

{Based on the bounds on the fluctuations we found, we might argue that, 
 fluctuations in 
gapped ground state obey a \emph{Gaussian statistics} (as they do in  non interacting theories).  A recent
proof of the Berry-Esseen theorem for the quantum case by Brand\~ao
\textit{et~al.}\cc{ref:Brandao2015-Berry-Esseen} hints that this
might be the case. A natural approach to this would be to tighten
our main theorem, replacing the exponential decay in the RHS of
inequality~\eqref{eq:LR1} by a Gaussian. }


Another intriguing direction to pursue is to incorporate LR, or one of its consequences, such as \Lem{lem:fluc} or
inequality \eqref{eq:fluctuations}, -- explicitly or implicitly --
in the construction of tensor networks in higher dimension (e.g.,
Projected entangled pair state, or PEPS\cc{verstraete2008matrix}).
By construction, these states satisfy the area-law, 
but we now know
that they should also satisfy local reversibility. This will speed
up the contraction of such tensor networks, which is the main
bottleneck in the variational algorithms\cc{ref:Cirac14-PEPS-alg,
ref:Cirac13-PEPS-contraction,anshu2016local,schwarz2016approximating}. 
A goal of paramount importance in
this context is to prove that PEPS are faithful representations of
gapped ground states. A good place to start studying this question
is in the 1D world. We know that MPS can describe both LR and non-LR
states (i.e., GHZ). The natural problem is then to pinpoint what is
needed for an MPS to describe an LR state.

Proving the area-law conjecture for gapped systems in 2D and
beyond remains a challenge. 
It would be interesting to see if
the additional structure of local reversibility of these states can
assist in such proofs, or at least provide new insights regarding
this important conjecture.
As a specific route, we  suggest to harness the LR \textit{in addition to the clustering},  to improve the upper bound by {Brand{\~a}o and Horodecki}\cc{Brandao}.

{Finally, it would be interesting to understand if local
reversibility could somehow be used to \emph{characterize} unique
gapped ground states.  In other
words, is local
reversibility also a sufficient condition for unique gapped ground
states? Strictly speaking, this is incorrect, as there are LR states
which are not gapped ground states. For example, the state
$\ket{000\cdots 0} + \epsilon(N)\ket{111\cdots 1}$ where
$\epsilon(N)$ decays faster than any polynomial is trivially LR, but
can never be a unique gapped ground state of $k$-local Hamiltonians
as long as $k \le N/2$ (see \Ref{ref:Facchi11-GHZ}). Nevertheless,
we may still ask if, in some sense, every LR state can be
approximated by a unique gapped ground state. If this is not the
case, it would be interesting to understand which are these LR
states that cannot be even approximated by gapped ground states.}

{Generalising our approach to mixed states and devising experimental protocols to measure local reversibility are important future challenge.}

\section*{Acknowledgement}
We are grateful to Naomichi Hatano, Tohru Koma, Hal Tasaki, Taku
Matsui, Tomoyuki Morimae, Kohtaro Kato and Dorit Aharonov for helpful discussions
and comments on related topics.  We also thank Naomichi
Hatano for valuable comments on the manuscript.  
This work was partially supported by the Program for Leading
Graduate Schools (Frontiers of Mathematical Sciences and Physics, or FMSP) MEXT Japan, 
and World Premier International Research Center Initiative (WPI) MEXT Japan.
TK also acknowledges the support from JSPS grant no.~2611111. 
Research at the Centre for Quantum Technologies is funded by the
Singapore Ministry of Education and the National Research
Foundation, also through the Tier 3 Grant random numbers from
quantum processes.

%
%
%



\begin{appendix}



%
%
%
%
%
%
%
%
%
%
%
\section{\bf Proof of Lemma~\ref{lem:macro}}
\label{proofmacro}
  Assume that $\ket{\psi}$ satisfies inequality~\eqref{def:LR}. Then for every integer $q>0$,
  there exists a $q$-local operator $R$ such that $R P_L
  \ket{\psi} = \ket{\psi} + \ket{\delta'}$, where $\norm{\delta'}^{2}\le
  \frac{\norm{P_L}}{|\bra{\psi}P_L\ket{\psi}|}
  f(q/\sqrt{|L|}) \le \frac{1}{|\alpha|^2}f(q/\sqrt{|L|})$. Therefore,
  \begin{align*}
    \beta\ket{\psi_b} &= \Id\ket{\psi} -P_L\ket{\psi}
      = (R-\Id)P_L\ket{\psi} - \ket{\delta'} \\
       &= \alpha(R-\Id)\ket{\psi_a} - \ket{\delta'}\,.
  \end{align*}
  By denoting $O = \alpha(R-\Id)/\beta$ and $\ket{\delta} =- \beta^{-1}\ket{\delta'}$, we have  $\ket{\psi_b} = O\ket{\psi_a} +\ket{\delta}$, where $\|\delta\|^{2} =\|\delta'\| ^{2}/|\beta| \le \frac{1}{|\alpha^2\beta|}f(q/\sqrt{|L|})$.
  This completes the proof of Lemma~\ref{lem:macro}.

\section{Proof of \Thm{thm:main}}
\label{sec:main-proof}

\subsection{Outline} \label{sec:outline_main_thm}
The proof of \Thm{thm:main} is rather technical, and therefore {we
first sketch it here, giving the full details in
the following section.}

Multiplying inequality~\eqref{eq:LR1} by
$|\bra{\gs}\Gamma_L\ket{\gs}|$, and writing for brevity
$\tilde{R}\EqDef \bra{\gs}\Gamma_L\ket{\gs}R$, we obtain
\begin{align}
  \norm{(\tilde{R}- \ket{\gs}\bra{\gs})\cdot \Gamma_L\ket{\gs}}
  \le  6\norm{\Gamma_L} e^{-2n_0 / \xi } \,.
\label{eq:AGSP}
\end{align}
So for the state to be LR, we need to find a $\tilde{R}$ whose
action on $\Gamma_L\ket{\gs}$ approximates the action of the ground
state projector $\ket{\gs}\bra{\gs}$ on it. In addition, in order to
satisfy the premise of the theorem, it has to be a $q$-local
operator. To this aim, we look for a low-degree polynomial $F_R(x)$
and write $\tilde{R}\EqDef F_R(H)$. Specifically, choosing a
polynomial of degree $n_0\EqDef \lfloor q/k \rfloor$ guarantees that
it will contain at most $q$-local terms, since, by definition, each
term in $H$ is $k$-local.

To understand the restrictions on $F_R(x)$ that
inequality~\eqref{eq:AGSP} poses, it is convenient to work in the
energy basis $\{\ket{E}\}$: expanding $\Gamma_L\ket{\gs} =\sum_E
c(E)\ket{E}$, we want \textit{i)} $F_R(0)=1$ (recall that have set
$E_0=0$), and \textit{ii)} $\left(\sum_{E\ge \delta E} |c(E)\cdot
F_R(E)|^2\right)^{1/2}\le 6\norm{\Gamma_L} e^{-2n_0/\xi}$. This is
achieved using two ideas, which are demonstrated in
\Fig{fig:schematic_picutre_proof_idea_LR}. 

The first idea is that the expansion of $\Gamma_L\ket{\gs}$ is
dominated by energies of at most $\orderof{|L|}$; beyond
that scale, $c(E)$ is exponentially decaying. This is a
direct corollary of Theorem~2.1 in \Ref{ref:Arad14-Edist}, which for
our case implies:
\begin{corol}[from Theorem~2.1 in \Ref{ref:Arad14-Edist}]
\label{cor:expE}
Let $\Pi^H_{\ge E}$ be the projector into the eigenspace of $H$ with
energies greater than or equal to $E$. Then
\begin{align}
\label{eq:expE}
  \sum_{E'\ge E} |c(E')|^2
   =\norm{\Pi^H_{\ge E}  \Gamma_L \ket{\gs}}^2 
   \le \norm{\Gamma_L}^2e^{-(E-2g|L|)/4gk} \,.
\end{align}
\end{corol}
In \Ref{ref:Arad14-Edist}, this theorem was proved under the more
restricted condition that every particle participates in at most $g$
interactions of norm 1, but this can be easily relaxed to the
current condition, given in definition~\eqref{def:H}.

The bound in~\eqref{eq:expE} implies that our polynomial should
mainly ``kill'' the energy excitations of $\Gamma_L\ket{\gs}$ in the
range $[\delta E, \orderof{|L|}]$. Following \Ref{ref:Arad-AL13}, we
let $F_R(x)$ be the $n_0$th order Chebyshev
polynomial\cc{ref:abramowitz1972handbook}, scaled such that
$x:[-1,1]\mapsto [\delta E, 2E_c+\delta E]$ and $F_R(0)=1$. As
discussed in the following section,
this polynomial fluctuates between $\pm e^{-2n_0/\xi}$ in the range
$[\delta E, 2E_c+\delta E]$, and then diverges like
$\orderof{(2x/E_c)^{n_0}}$. It is our choice of $E_c$ in
\Thm{thm:main} which guarantees that this divergence is cancelled by
the exponential decay of Corollary~\ref{cor:expE}.  After a rather
straightforward calculation, one can show that total contributions
of the energy segments $[\delta E, 2E_c+\delta E]$ and $[2E_c+\delta
E, \infty)$ to
$\norm{(\tilde{R}-\ket{\gs}\bra{\gs})\cdot\Gamma_L\ket{\gs}}$ is
exponentially small.

\subsection{Full proof}

Following the proof's sketch in the previous section of the main text of the paper, we start from
inequality~\eqref{eq:expE}. Our goal is to find a
polynomial $F_R(x)$ such that the action of the operator
$\tilde{R}\EqDef F_R(H)$ on the state $\Gamma_L\ket{\gs}$
approximates the action of the ground state projector
$\ket{\gs}\bra{\gs}$ on it. As $H$ is a $k$-local operator, choosing
$n_0\EqDef \lfloor q/k \rfloor$ guarantees that $\tilde{R}$ is a
$q$-local operator.

Working in the eigenbasis of $H$, we expand $\Gamma_L\ket{\gs} =
\sum_{E} c(E)\ket{E}$, and as $F_R(H)$ is diagonal in this basis,
\begin{align*}
  [F_R(H)-\ket{\gs}\bra{\gs}]\Gamma_L\ket{\gs} &=
   (F_R(0)-1)c(0)\ket{\gs} + \sum_{E\ge\delta E} F_R(E)c(E)\ket{E} \,.
\end{align*}
Therefore, for inequality $(12)$ to hold, it is sufficient
that
\begin{align}
\label{eq:cond1}
  &F_R(0) =1 \\
\label{eq:cond2}
  &\left(\sum_{E>\delta E} |c(E) F_R(E)|^2\right)^{1/2}
    \le 6\norm{\Gamma_L}e^{-2n_0/\xi} \,. 
\end{align}
As noted in the outline of the proof in the previous section, to
prove these properties we use two ideas. The first is that the
weight of the high energy excitations in $\Gamma_L\ket{\gs}$ decays
exponentially, as shown in Corollary~\ref{cor:expE} of Section~\ref{sec:outline_main_thm}. The second is to take
$F_R(x)$ to be a scaled version of the $n_0$'th order Chebyshev
polynomial. Let us start from the second idea. The $n$th order
Chebyshev polynomial\cc{ref:abramowitz1972handbook} of the first
kind is given by
\begin{align}
\label{def:Cheby}
  T_n(x)\EqDef \frac{(x + \sqrt{x^2 -1})^n 
    +(x - \sqrt{x^2-1})^n}{2} \,.
\end{align} 
Equivalently, for $x\in[-1,1]$ it is given by
$T_n(x)=\cos(n\arccos(x))$, and for $|x|> 1$ by
$T_n(x)=\cosh(n\arccosh(x))$. What makes the Chebyshev polynomial so
useful to our purpose are the properties that are summarized in the
following lemma, whose proof is given in \Sec{sec:Cheby-proof}:
\begin{lemma}
\label{lem:Cheby}
  \begin{align}
  \label{eq:prop1}
    |T_n(x)|&\le 1 \ , &\text{for $|x|\le 1$} \\
  \label{eq:prop2}
    |T_n(x)| &\le \frac{1}{2} (2|x|)^n \ , &\text{for $|x|\ge 1$} \\
  \label{eq:prop3}
    |T_n(x)| &\ge \frac{1}{2}\exp{
      \left(2n\sqrt{\frac{|x|-1}{|x|+1}}\right)} \ , 
        &\text{for $|x|\ge 1$}
  \end{align}
\end{lemma}
Setting
\begin{align}
  \xi \EqDef \sqrt{1+\frac{2E_c}{\delta E}}, \quad
  \text{and}\quad
  E_c\EqDef  g|L| +8gkn_0 \,, 
  \label{def:Ec2}
\end{align}
we define $F_R(x)$ to be the polynomial
\begin{align}
  F_R(x) \EqDef \frac{T_{n_0}(\frac{x-\delta E}{E_c}-1)}
    {T_{n_0}(\frac{-\delta E}{E_c}-1)} \,.
\label{def:Fr}
\end{align}
In other words, we defined it to be the $n_0$th order Chebyshev
polynomial, scaled such that $x: [-1,1] \mapsto [\delta E,
2E_c+\delta E]$ and $F_R(0)=1$. Clearly, this definition satisfies
\Eq{eq:cond1}. Let us see why it also satisfies
inequality~\eqref{eq:cond2}.

We begin by applying \Lem{lem:Cheby} to the definition
of $F_R(x)$, which implies that for $\delta E \le x\le 2E_c+ \delta
E$,
\begin{align}
  |F_R(x)| \le 2 e^{-2n_0 /\xi}\,,
\label{eq:Upper_bound_Fr_x1ee}
\end{align}
and for $x\ge 2E_c+ \delta E$,
\begin{align}
  |F_R(x)| &\le \biggl(\frac{2x-2\delta E}{E_c}-2\biggr)^{n_0}
    e^{-2n_0 /\xi } \,.
\label{eq:Upper_bound_Fr_x2ee}
\end{align}
For brevity, we define the low and high energy ranges $I_{\rm LOW}\EqDef
[\delta E, 2E_c+\delta E)$ and $I_{\rm HI}\EqDef [2E_c+\delta E,
\infty)$. Then using the triangle inequality, we split the sum in
the LHS of \eqref{eq:cond2}
\begin{align*}
  &\left(\sum_{E>\delta E} |c(E) F_R(E)|^2\right)^{1/2}
  \le \left(\sum_{E\in I_{\rm LOW}}
    |c(E) F_R(E)|^2\right)^{1/2} + \left(\sum_{E\in I_{\rm HI}}
      |c(E) F_R(E)|^2\right)^{1/2} \,,
\end{align*}  
and bound each term separately. The low-energy term is bounded by
\begin{align}
  2e^{-2n_0/\xi} \left(\sum_{E\in I_{\rm LOW}}
    |c(E)|^2\right)^{1/2}
  \le 2e^{-2n_0/\xi} \norm{\Gamma_L\ket{\gs}}
  \le 2\|\Gamma_L\| e^{-2n_0 /\xi} \,,
\label{eq:Chebyshev_Vanishing_higher_energy_eigenstates_1}
\end{align}
which follows from Inequality~\eqref{eq:Upper_bound_Fr_x1ee} and the
fact that $\sum_{E\in I_{\rm LOW}}|c(E)|^2\le \sum_E |c(E)|^2 = \norm{\Gamma_L\ket{\gs}}^2$.

To finish the proof, we will show that the high energies term is
upper bounded by $4\|\Gamma_L\| e^{-2n_0 /\xi}$. To this aim, we
write $I_{\rm HI} = I_1 \cup I_2 \cup I_3 \cup\ldots$, where 
$I_j\EqDef [2E_c+\delta E+(j-1)\eta, 2E_c+\delta E +
j\eta)$ and $\eta$ is a positive constant
which will be set afterward. Using the triangle inequality once
more, we get
\begin{align*}
  \left(\sum_{E\in I_{\rm HI}}
      |c(E) F_R(E)|^2\right)^{1/2} \le
  \sum_{j=1}^\infty \left(\sum_{E\in I_j} 
    |c(E) F_R(E)|^2\right)^{1/2} \,.
\end{align*}
Clearly, for each $I_j$ segment
\begin{align*}
  \left(\sum_{E\in I_j} |c(E) F_R(E)|^2\right)^{1/2}
   &\le \max_{x\in I_j}|F_R(x)|
   \left(\sum_{E\in I_j} |c(E)|^2\right)^{1/2} \,.
\end{align*}
As $|F_R(x)|$ monotonically increases for $x\ge
2E_c+\delta E$ (which follows from the fact that the Chebyshev
polynomial is monotonic for $x\ge 1$), it follows that
\begin{align*}
  \max_{x\in I_j}|F_R(x)|\le |F_R(2E_c+\delta E+j\eta)| \,.
\end{align*}
To bound the other term, we use Corollary~\ref{cor:expE},
which gives us
\begin{align*}
  \left(\sum_{E\in I_j} |c(E)|^2\right)^{1/2}
  \le \left(\sum_{E\ge 2E_c+\delta E+(j-1)\eta} 
    |c(E)|^2\right)^{1/2} 
   \le  \norm{\Gamma_L}
     e^{-\lambda(2E_c+\delta E + (j-1) \eta-2g |L|))} \,,
\end{align*}
where we have defined
\begin{align}
\label{def:lambda}
  \lambda\EqDef \frac{1}{4gk}  \,.
\end{align}
Together, this gives us
\begin{align*}
  &\left(\sum_{E\in I_j} |c(E) F_R(E)|^2\right)^{1/2}
   \le \norm{\Gamma_L} e^{\lambda\eta} \cdot \left |F_R(2E_c+\delta E+j\eta)\right|  
     e^{-\lambda(2E_c+\delta E + j\eta-2g |L|)} \,.
\end{align*}
The final step is to show that for $x\ge 2E_c+ \delta E$,
\begin{align}
  |F_R(x)|\cdot e^{-\lambda(x -2g |L|)}  
    \le e^{-2n_0 /\xi} \cdot e^{-\lambda(x -2g |L|)/2}   
\label{Chebyshev_F_R_H_exp2}
\end{align}
(see Subsection~\ref{sec:exp-ineq} for a proof), which leads to
\begin{align*}
  &\left(\sum_{E\in I_j} |c(E) F_R(E)|^2\right)^{1/2}
  \le \norm{\Gamma_L}   e^{-2n_0 /\xi} e^{\lambda\eta}\cdot e^{-\lambda(2E_c+\delta E + j\eta-6g |L|)/2 } \,.
     \nonumber
\end{align*}

Summing over all $j\ge 1$, then gives us
\begin{align*}
  &\left(\sum_{E\in I_{\rm HI}}|c(E) F_R(E)|^2\right)^{1/2}
  \le \norm{\Gamma_L}e^{-2n_0/\xi} \cdot   e^{-\lambda(2E_c + \delta E -2g|L|)/2}
    \cdot e^{\lambda\eta} \sum_{j=1}^\infty e^{-j\eta\lambda/2} \,.
\end{align*}
Using the definition of $E_c$ in \Eq{def:Ec2}, we find that
$e^{-\lambda(2E_c + \delta E -2g|L|)/2} = e^{-\lambda(16gkn_0+\delta
E)/2} \le 1$, and calculating the geometrical sum we get
$e^{\lambda\eta} \sum_{j=1}^\infty e^{-j\eta\lambda/2} =
e^{\lambda\eta/2}/(1-e^{-\lambda\eta/2})$, which can be minimized to
$4$ by choosing $\eta$ such that $e^{\lambda\eta/2}=2$. All
together, we therefore get
\begin{align}
  \left(\sum_{E\in I_{\rm HI}}|c(E) F_R(E)|^2\right)^{1/2} 
  \le 4\norm{\Gamma_L}e^{-2n_0/\xi} \,,
\end{align}
which completes the proof.

\subsubsection{Proof of \Lem{lem:Cheby}}
\label{sec:Cheby-proof}

\begin{proof}
  Inequality~\eqref{eq:prop1} follows directly from the identity
  $T_n(x)=\cos(n\arccos(x))$, which is valid for $|x|\le 1$. For the
  other inequalities, first note that $T_n(-x) = (-1)^nT_n(x)$, 
  which implies $|T_n(x)| = |T_n(|x|)|$, and so it is sufficient to
  prove inequalities~(\ref{eq:prop2},~\ref{eq:prop3}) for $x>1$. 
  
  To prove inequality~(\ref{eq:prop2}), consider the general inequality
  \begin{align}
    (2 x -y)^n +y^n \le (2x)^n \,, \label{eq:Chebyshev-upper-proof1}
  \end{align}
  which is valid for any $x\ge 1$ and $0\le y\le 1$ (the inequality
  can be proved by differentiating $(2x)^n- \big((2 x -y)^n
  +y^n\big)$ with respect to $x$, and noting for $x\ge 1$ and $0\le
  y\le 1$ it is a monotonically increasing function of $x$, and its 
  minimum value 0, which is obtained for $x=1$ and $y=0$).
  Choosing $y=x-\sqrt{x^2 -1}$, the LHS of inequality
  \eqref{eq:Chebyshev-upper-proof1} becomes $2T_n(x)$, which proves
  \eqref{eq:prop2}.
  
  For inequality~\eqref{eq:prop3}, we set $t\EqDef \arccosh(x)$, and
  then by the identity $T_n(x)=\cosh(n\arccosh(x))$, we conclude that
  for $x>1$,
  \begin{align*}
    T_n(x) = \cosh(nt) = \frac{1}{2}\left(e^{nt} + e^{-nt}\right) 
      \ge \frac{1}{2}e^{nt} \,.
  \end{align*}
  To finish the proof, we need to show that for $x>1$, $t\ge
  2\sqrt{\frac{x-1}{x+1}}$. This follows from the fact that $t/2\ge
  \tanh(t/2)$, and the trigonometric identity $\tanh(t/2) =
  \sqrt{\frac{\cosh(t)-1}{\cosh(t)+1}}$.
\end{proof}

\subsubsection{Derivation of the inequality~(\ref{Chebyshev_F_R_H_exp2})}
\label{sec:exp-ineq}

From  inequality~\eqref{eq:Upper_bound_Fr_x2ee}, we have
\begin{align*}
  &|F_R(x)|
    \le e^{-2n_0 /\xi} 
      \biggl(\frac{2x-2\delta E}{E_c}-2\biggr)^{n_0} \,,
\end{align*}
for $x\ge 2E_c+ \delta E$. To prove
inequality~\eqref{Chebyshev_F_R_H_exp2}, we will show that
$\bigl[(2x-2\delta E)/E_c-2\bigr]^{n_0}e^{-\lambda (x -6g |L|)/2}\le
1$ for $x\ge 2E_c+ \delta E$, or, equivalently, that its logarithm 
\begin{align*}
  G(x)\EqDef -\frac{\lambda}{2}(x- 6g |L|) 
    + n_0 \log \Bigl (\frac{2x-2\delta E}{E_c}-2\Bigr)
\end{align*}
is negative. This follows from the facts that
\begin{align*}
  G(2E_c+\delta E) = -2n_0 -\frac{\lambda \delta E}{2}+  n_0 \log 2 
    < 0 \,,
\end{align*}
and for every $x\ge 2E_c+\delta E$,
\begin{align*}
  \frac{dG(x)}{dx} = -\frac{\lambda}{2} 
    + \frac{n_0}{x-E_c-\delta E}
  \le -\frac{\lambda}{2} + \frac{n_0}{E_c} 
  = -\frac{\lambda}{2}
    +\frac{\lambda}{2+\frac{3g \lambda |L|}{n_0}} < 0 \,.
\end{align*}


{
\section{Difference between degenerate and non-degenerate topological orders}
\label{sec:toric}}

{
{
In the case of the toric code model, we find that the LR depends on the topology of the ambient manifold: 
LR holds on a sphere but is violated on  non simply connected geometries (implying a non trivial ground-manifold). 
It is  well-known, however, that the topological entanglement entropy is non-vanishing for toric code model ground states living in lattice with any topology~\cite{topo_ent_kit,topo_ent_wen}. 
Indeed, the difference  between the two kind of ground states can be resolved in terms of the irreducible multiparty correlation.}}

 \begin{figure}[tt]
\centering
\includegraphics[clip, scale=0.6]{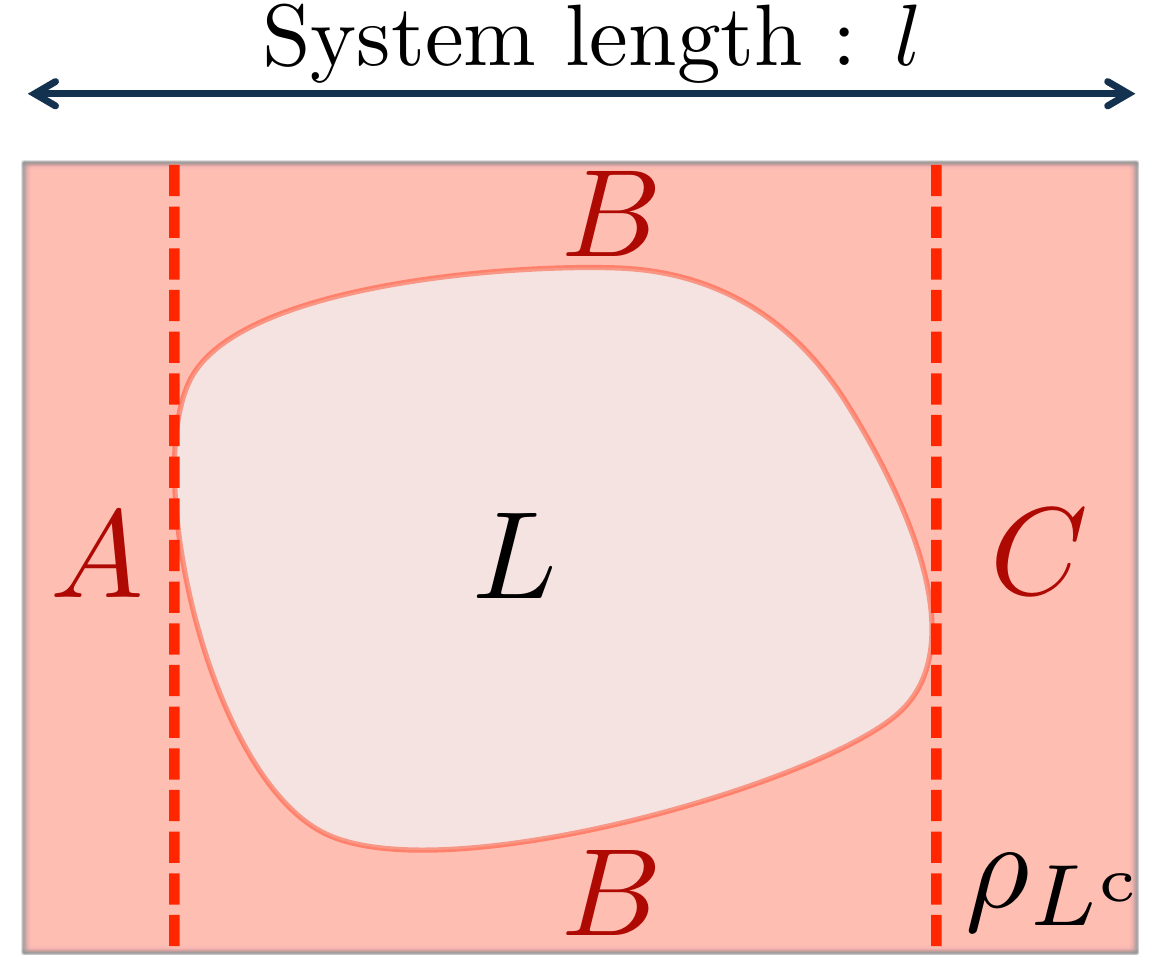}
\caption{Multipartite correlations in the surface code. 
In the ground states of the Kitaev model on sphere, it has no multi-party correlation (or contains only low-degree of correlations), but  
collective properties of the low-degree of correlations induce multi-party correlation when we look at reduced region of the system, say $L^{\co}$. 
Indeed, if we split the region $L^{\co}$ into $A$, $B$ and $C$, we obtain non-trivial value of the topological entanglement entropy.
}
\label{fig:Surface}
\end{figure}
{
{The notion of irreducible multipartite correlation has been first introduced  in 
Ref.~\cite{Linden_irred} to characterize the multipartite correlations in a quantum state. 
It was noted recently that  such notion is  equivalent to the topological entanglement entropy if the state has zero-correlation length~\cite{Kato:cor}.
As explained in Refs~\cite{irred:Zhou,liu2014irreducible},
we have two kinds of multipartite correlation, which we refer to as `effective multiparty correlations', distinct from `inherent multipartite correlations.' 
The topological entanglement entropy cannot distinguish them. 
We have:}}

{~}
{
{{\it i) The degenerate topological order, as that one of the  toric code on a torus, has genuine multiparty correlation of the `inherent' type involving  $\mathcal{O}(l)$  spins ($l$: the system length).}}}
 
{~}
{ 
{{\it ii) The non-degenerate topological order, as the toric code on the sphere,  has low degree of inherent multiparty correlations involving  $\mathcal{O}(1)$  spins, but have  the `effective' type involving  $\mathcal{O}(l)$ spins. }}
}
{~}

{In other words, a non-vanishing  topological entanglement entropy in non-degenerate topological order arises just because of such multiparty low-degree correlations. 
There, we have no high-degree multiparty correlations if we look at the \textit{total system}; in contrast, multiparty correlations of $\mathcal{O}(l)$ can be effectively induced
by  tracing  out some finite suregions (See Fig.~\ref{fig:Surface})~\cite{liu2014irreducible}.
Such a conditional many-body correlations can appear in short-range entangled state~\cite{Bavyi_unpublished,Haar_slide} or even in classical models~\cite{PhysRevE.85.046209}.}

{
In this way, we can see qualitative difference between the degenerate and the non-degenerate topological orders in terms of the irreducible multiparty correlation, which results in LR of the surface code and non-LR of the toric code. 
Being our approach able to detect such `fine structure' in the nature of the multipartite correlations, the LR tells degenerate topological order apart  from non-degenerate topological order.
}

\section{Symmetry-restricted Local Reversibility}
\label{srLR}
Symmetry restricted LR states (SRL) can be introduced along very
similar lines used in \Sec{section:LR}.  Let's consider a given
Hamiltonian $H$ enjoying a global symmetry $G$; let $\ket{\psi}$ be
the ground state of $H$. We say that the state $\ket{\psi}$ is SLR
iff the property (\ref{def:LR}) holds with a $q$-local operator $R$
enjoying the same symmetry group of the Hamiltonian: $[R,G]=0$.

Here we present an example of states which are \emph{not}
SLR.  Cluster states provide an example of
SPTO.  The 1D cluster states\cc{son2011quantum} are the ground
states of the Hamiltonian
\begin{align}
  H_C=\sum_{i=1}^L \sigma_{i-1}^x\sigma_i^z \sigma_{i+1}^x \,,
\end{align}
which enjoys a global symmetry $Z_2\times
Z_2$\cc{son2012topological}.  With the boundary conditions
$\sigma_0^x=\sigma_{L+1}^x=\Id$, the ground space of $H_C$ is unique
with a spectral gap. For $\sigma_0^x=\sigma_{L+1}^x=0$, in contrast,
the ground space is four-fold degenerate because the two stabilizers
(out of $L$) $\sigma_{0}^x\sigma_1^z \sigma_{2}^x$ and
$\sigma_{L-1}^x\sigma_L^z \sigma_{L+1}^x$ can be fixed at
will\cc{son2012topological}. Let \{$\ket{\gs_\alpha},
\alpha=0,1,2,3\}$ be spanning the ground state manifold.  Due to the
symmetry-protected topological order of the system, it follows that
the ground states $\ket{\gs_\alpha}$ cannot be distinguished by any
local operator $o_X$ in $Z_2\times Z_2$:
\begin{align}
  \bra{\gs_\alpha}o_X\ket{\gs_\alpha}
    =\bra{\gs_\beta}o_X\ket{\gs_\beta},
    \quad {\rm and} \quad \bra{\gs_\alpha}o_X\ket{\gs_\beta}=0.  
    \label{Topo_basic}
\end{align}
with $|X| \le cN$ ($c=\orderof{1}$). Using these conditions, the
symmetry-restricted non-LR of $\ket{\gs_\alpha}$ follows from the
same arguments that were used in the proof of the non-LR of the
toric code.

\section{Critical exponents}
\label{sec:critical}

Here, we  derive inequality~\eqref{eq:critical-exp} for the critical
exponents $z$, $\eta$, $\gamma$ and $\nu$ under the scaling
ansatz~\eqref{Scaling_ansatz_flucutation}\cc{ref:Vojta03-Qphase-trans,
ref:Conti94-Quantum-scaling}. Recall
that we are considering a local Hamiltonian system at $T=0$ which is
driven towards critically, and let $A=\sum_i a_i$, where $a_i$ are
single particle operators that correspond to a local order parameter
(e.g., spin localized at site $i$ leading to the magnetization along a given axes). Our starting point is
inequality~\eqref{eq:fluctuations}, namely
\begin{align}
  \delta E \cdot (\Delta A)^2 \le {\rm const} \cdot N\,.
  \label{trade_off_fluctuation_spectral_gap_s}
\end{align}

We first define the variance $(\Delta A_t)^2$ which depends on 
time as $(\Delta A_t)^2 \EqDef \big\langle \big( A(t) - \langle A
\rangle\big)\cdot\big(\langle A - \langle A\rangle\big)\big\rangle$,
where $A(t)= e^{-i H t} A e^{i H t}$.  The variance $(\Delta A_t)^2$
reduces to the summation of the correlation functions:
\begin{align} 
  (\Delta A_t )^2&=\sum_{i,j=1}^N  
     \langle a_i(t)a_j  \rangle 
    - \langle a_i \rangle  \langle a_j\rangle
  \EqDef \sum_{i,j=1}^N C_{i,j}(t), \notag 
\end{align}
where $a_i(t)\EqDef e^{-i H t} a_i e^{i H t}$ for $i=1,2,\ldots N$.
Note that $(\Delta A_{t=0} )^2$ is equal to $(\Delta A)^2=\langle
A^2 \rangle - \langle A\rangle^2$. In the following, we denote
$C_{i,j}(t)=C({\bm r},t)$ under the assumption of the translation
symmetry.

Now, we adopt the following scaling
ansatz\cc{ref:Conti94-Quantum-scaling}:
\begin{align} 
  S({\bm q},\omega; \xi) = \xi^{2-\eta} D({\bm q}\xi,\omega \xi^z),  
\label{Scaling_ansatz_flucutation}
\end{align}
where $\xi$ is the correlation length and $S({\bm q},\omega;\xi)$ is
the spatial-temporal Fourier component of $C({\bm r},t)$, namely
\begin{align} 
  S({\bm q},\omega;\xi) &=\int_{{\bm r}} 
    \int_{t}C({\bm r},t) e^{-i({\bm q}\cdot{\bm r}+\omega t)} 
      d{\bm r}dt.
\end{align}
We also define $S({\bm q};\xi)$ as
\begin{align} 
  S({\bm q};\xi)=\frac{1}{2\pi}\int_{-\infty}^{\infty} 
    S({\bm q},\omega;\xi)  d\omega.
\end{align}
We can see that the static fluctuation $(\Delta A_{t=0})^2$ is equal
to $N S({\bm q}=0;\xi)$ by expanding $S({\bm q}=0;\xi)$.

We then obtain the scaling of $S({\bm q}=0;\xi)\propto
\xi^{2-\eta-z}$ by taking the
scaling~\eqref{Scaling_ansatz_flucutation} for $S({\bm
q},\omega;\xi)$, and hence we have $(\Delta A_{t=0})^2/N \propto
\xi^{2-\eta-z}$. We also have the scaling of the energy gap as
$\delta E_0 \propto \xi^{-z}$\cc{ref:Conti94-Quantum-scaling} by the use of the
dynamical critical exponent~$z$. At a critical point, where the
correlation length is as large as the system length, the
inequality~\eqref{trade_off_fluctuation_spectral_gap_s} reduces to
\begin{align} 
  -z\le -(2-\eta-z)
\end{align}
in the infinite volume limit ($N\to\infty$).  This reduces to the
inequality~\eqref{eq:critical-exp} in the main manuscript.

We close the section applying inequality~\eqref{eq:fluctuations} to a system with long-range interactions:  the  Lipkin-Meshcov-Glick model $H_{\rm
LMG}=-\frac{\lambda}{N}\sum_{i<j}(\sigma_i^x\sigma_{j}^x +\gamma
\sigma_i^y \sigma_{j}^y) +\sum_{i=1}^{N} h \sigma_i^x $ with
$|\gamma|\le 1$. At the critical point $\lambda=|h|$, we have the
scaling\cc{dusuel2004finite} of $\delta E \propto N^{-1/3}$ and
$(\Delta M_x)^2\propto N^{4/3}$, where $M_x$ is the magnetization in
the $x$ direction, $M_x= \sum_{i=1}^{N} \sigma_i^x $. Thus, the
spectral gap and the fluctuation can give the non-trivial sharp
upper bounds to each other. 

\section{The quality of the mean-field approximation}
\label{sec:MF}

Let $\ket{\gs}$ be the unique ground state of a gapped local
Hamiltonian, and let $\rho_{ij}, \rho_i, \rho_j$ be its
two-particles and one-particles reduced density matrices. We want to
estimate the error of the mean-field approximation $\rho_{ij} \to
\rho_i\otimes\rho_j$ by proving inequality~\eqref{eq:meanF} in the main text.
For simplicity, we set $i=1$ and show that
\begin{align}
\label{eq:total-diff}
  \sum_{j \in L} \norm{\rho_{1,j}-\rho_{1}\otimes \rho_{j}}
    \le {\rm const} \cdot \sqrt{|L|/\delta E}\,.
\end{align}
First, note that we can always find a set of $d^2$ projectors
$\{P_1^{(m)}\}$ onto the spin $i=1$ that satisfy
\begin{align}
  \norm{\rho_{1,j}-\rho_{1}\otimes \rho_{j}} 
    \le \sum_{m=1}^{d^2} \norm{P_1^{(m)}
      (\rho_{1,j}-\rho_{1}\otimes\rho_{j})P_1^{(m)}} \,, 
\label{eq:Decomp_Proj_Sum_Op}
\end{align}
where $d$ is the local spin dimension. For example, in
the case of spin-1/2 systems ($d=2$), we can take $P_1^{(1)} =
\ket{0_1}\bra{0_1}$, $P_1^{(2)} = \ket{1_1}\bra{1_1}$, $P_1^{(3)} =
\ket{+_1}\bra{+_1}$, $P_1^{(4)} = \ket{-_1}\bra{-_1}$, with $\ket{\pm_1}\EqDef (\ket{0_1} \pm \ket{1_1})/\sqrt{2}$. Indeed, defining
$\delta \rho_{1,j}\EqDef \rho_{1,j}-\rho_{1}\otimes \rho_{j}$, we
get
\begin{align*}
  \norm{\delta \rho_{1,j}} 
    &\le \norm{\bra{0_1}\delta \rho_{1,j}\ket{0_1}} 
    + \norm{\bra{1_1}\delta \rho_{1,j} \ket{1_1}} 
    + \norm{\bra{0_1}\delta \rho_{1,j} \ket{1_1} 
      +\bra{1_1}\delta \rho_{1,j} \ket{0_1}}\nonumber \\
  &= \norm{\bra{0_1}\delta \rho_{1,j} \ket{0_1}} + 
      \norm{\bra{1_1}\delta \rho_{1,j} \ket{1_1}} 
  + \norm{\bra{+_1}\delta \rho_{1,j} \ket{+_1} 
      - \bra{-_1}\delta \rho_{1,j} \ket{-_1}}\nonumber \\
  &\le \norm{\bra{0_1}\delta \rho_{1,j} \ket{0_1}} 
    + \norm{\bra{1_1}\delta \rho_{1,j} \ket{1_1}} 
  + \norm{\bra{+_1}\delta \rho_{1,j} \ket{+_1}} 
    + \norm{\bra{-_1}\delta \rho_{1,j} \ket{-_1}} \,.
\end{align*}
The proof for higher $d$ follows the same lines.

Summing inequality~\eqref{eq:Decomp_Proj_Sum_Op} over all $j\in L$
gives 
\begin{align*}
  \sum_{j\in L} \norm{\rho_{1,j}-\rho_{1}\otimes \rho_{j}} 
    \le \sum_{m=1}^{d^2} 
      \sum_{j\in L} \norm{P_1^{(m)} 
        (\rho_{1,j} - \rho_{1}\otimes\rho_{j})P_1^{(m)}}\,.
\end{align*}
To prove inequality~\eqref{eq:total-diff}, we will show an upper
bound of $\sum_{j\in L}\norm{P_1^{(m)} (\rho_{1,j}-\rho_{1}\otimes
\rho_{j} )P_1^{(m)}}$ for arbitrary $m$.  

Defining $\rho_{j}^{(m)} \EqDef \Tr_1 ( P_1^{(m)}\rho_{1,j}
P_1^{(m)} )$, where $\Tr_i (\cdots)$ is the partial trace over the
$i$th spin, we get
\begin{align*}
  P_1^{(m)} (\rho_{1,j}-\rho_{1}\otimes \rho_{j} )P_1^{(m)} = P_1^{(m)}\otimes \bigl(\rho_{j}^{(m)} 
      - \bra{\gs}P_1^{(m)}\ket{\gs}\cdot\rho_{j}\bigr) \,.
\end{align*}
Clearly, $\norm{P_1^{(m)}\otimes \bigl(\rho_{j}^{(m)} -
\bra{\gs}P_1^{(m)}\ket{\gs}\cdot\rho_{j}\bigr)} =
\norm{\rho_{j}^{(m)} - \bra{\gs}P_1^{(m)}\ket{\gs}\cdot\rho_{j}}$.
Moreover, there always exists a rank-1 projector $P_j^{(m)}$ such
that 
\begin{align*}
  \norm{\rho_{j}^{(m)} - \bra{\gs}P_1^{(m)}\ket{\gs}\cdot\rho_{j}}
   = s_j^{(m)} \cdot \Tr \Bigl[ P_j^{(m)} \bigl(\rho_{j}^{(m)} - \bra{\gs} P_1^{(m)}\ket{\gs} \cdot \rho_{j} \bigr )\Bigr] \,,
\end{align*}
where $s_j^{(m)} \EqDef \textrm{sign}\Bigl\{ \Tr \bigl[ P_j^{(m)} (
\rho_{j}^{(m)} - \bra{\gs} P_1^{(m)}\ket{\gs}\cdot \rho_{j} )
\bigr] \Bigr\}$. Therefore,
\begin{align*}
  \norm{P_1^{(m)} (\rho_{1,j}-\rho_{1}\otimes \rho_{j} )P_1^{(m)}}
  &= s_j^{(m)} \cdot \Tr \Bigl[ P_j^{(m)} \bigl(\rho_{j}^{(m)} 
     - \bra{\gs} P_1^{(m)}\ket{\gs} \cdot \rho_{j} \bigr )\Bigr] \\
  &= s_j^{(m)}\cdot\bigl[\bra{\gs} P_1^{(m)} P_j^{(m)} \ket{\gs}
     - \bra{\gs} P_1^{(m)}\ket{\gs} 
       \cdot\bra{\gs} P_j^{(m)}\ket{\gs}\bigr] \,.
\end{align*}
We now define the additive operator
\begin{align*}
  A^{(m)} \EqDef \sum_{j\in L} s_j ^{(m)}\cdot P_j^{(m)} \,.
\end{align*}
Then from the above calculation,
\begin{align*}
  \sum_{j\in L} \norm{P_1^{(m)} (\rho_{1,j}
    -\rho_{1}\otimes \rho_{j})P_1^{(m)}} = \bra{\gs} P_1^{(m)} A^{(m)} \ket{\gs} 
     - \bra{\gs} P_1^{(m)}\ket{\gs} 
       \cdot\bra{\gs}A^{(m)}\ket{\gs} \,.
\end{align*}
But
\begin{align*}
  \bra{\gs} P_1^{(m)} A^{(m)} \ket{\gs} 
     - \bra{\gs} P_1^{(m)}\ket{\gs} 
       \cdot\bra{\gs}A^{(m)}\ket{\gs}
  &= \bra{\gs} P_1^{(m)} \cdot 
  \Big[ A^{(m)}\ket{\gs}-\bra{\gs}A^{(m)}\ket{\gs}\ket{\gs}\Big] \\
  &\le \norm{P^{(m)}_1\ket{\gs}} \cdot 
   \norm{A^{(m)}\ket{\gs} -\bra{\gs}A^{(m)}\ket{\gs}\ket{\gs}} \,,
\end{align*}
and as $\norm{A^{(m)}\ket{\gs}
-\bra{\gs}A^{(m)}\ket{\gs}\ket{\gs}} = \Delta A^{(m)}$, we
conclude that
\begin{align*}
  \sum_{j\in L} \norm{P_1^{(m)} (\rho_{1,j}
    - \rho_{1}\otimes \rho_{j} )P_1^{(m)}} 
    &\le \Delta A^{(m)} 
    \le {\rm const} \cdot \sqrt{|L|/\delta E}
    \,.
\end{align*}
Here, the last inequality comes from the
inequality~\eqref{eq:fluctuations} in the main text, which applies
in this case since $A^{(m)}$ is an additive operator on $L$.
Combining this with inequality~\eqref{eq:Decomp_Proj_Sum_Op}
completes the proof.

{~}

\noindent\textbf{Optimality of the bound} \\
When $\delta E=\orderof{1}$, inequality~\eqref{eq:total-diff}
reduces to
\begin{align}
  \sum_{j\in L} \norm{\rho_{1,j}-\rho_{1}\otimes \rho_{j}}  
    \le {\rm const} \cdot \sqrt{|L|} \,. 
\end{align}
We can ensure that this upper bound is qualitatively optimal by
considering the state
\begin{align}
  \frac{1}{\sqrt{2}}\ket{0_1}\ket{0_20_3\cdots 0_N} 
    + \frac{1}{\sqrt{2}}\ket{1_1}\ket{{\rm W}_{2,\ldots,N} } \,, 
\label{optimal_monogamy}
\end{align}
where $\ket{{\rm W}_{2,\ldots,N}}$ is the W state for the spins $2,
3, \ldots,N$. 
We note that this state satisfies inequality $(\Delta A)^2 \le \orderof{|L|}$ \cc{ref:Shimizu05-macroE}, which is equivalent to the inequality~\eqref{eq:fluctuations} in the case of $\delta E=\orderof{1}$.
Interestingly, the state in~\eqref{optimal_monogamy} also gives the
upper limit of the monogamy inequality of the
entanglement~\cite{ref:Osborne06-Monogamy}.

\end{appendix}


\begin{thebibliography}{65}%
\makeatletter
\providecommand \@ifxundefined [1]{%
 \@ifx{#1\undefined}
}%
\providecommand \@ifnum [1]{%
 \ifnum #1\expandafter \@firstoftwo
 \else \expandafter \@secondoftwo
 \fi
}%
\providecommand \@ifx [1]{%
 \ifx #1\expandafter \@firstoftwo
 \else \expandafter \@secondoftwo
 \fi
}%
\providecommand \natexlab [1]{#1}%
\providecommand \enquote  [1]{``#1''}%
\providecommand \bibnamefont  [1]{#1}%
\providecommand \bibfnamefont [1]{#1}%
\providecommand \citenamefont [1]{#1}%
\providecommand \href@noop [0]{\@secondoftwo}%
\providecommand \href [0]{\begingroup \@sanitize@url \@href}%
\providecommand \@href[1]{\@@startlink{#1}\@@href}%
\providecommand \@@href[1]{\endgroup#1\@@endlink}%
\providecommand \@sanitize@url [0]{\catcode `\\12\catcode `\$12\catcode
  `\&12\catcode `\#12\catcode `\^12\catcode `\_12\catcode `\%12\relax}%
\providecommand \@@startlink[1]{}%
\providecommand \@@endlink[0]{}%
\providecommand \url  [0]{\begingroup\@sanitize@url \@url }%
\providecommand \@url [1]{\endgroup\@href {#1}{\urlprefix }}%
\providecommand \urlprefix  [0]{URL }%
\providecommand \Eprint [0]{\href }%
\providecommand \doibase [0]{http://dx.doi.org/}%
\providecommand \selectlanguage [0]{\@gobble}%
\providecommand \bibinfo  [0]{\@secondoftwo}%
\providecommand \bibfield  [0]{\@secondoftwo}%
\providecommand \translation [1]{[#1]}%
\providecommand \BibitemOpen [0]{}%
\providecommand \bibitemStop [0]{}%
\providecommand \bibitemNoStop [0]{.\EOS\space}%
\providecommand \EOS [0]{\spacefactor3000\relax}%
\providecommand \BibitemShut  [1]{\csname bibitem#1\endcsname}%
\let\auto@bib@innerbib\@empty
\bibitem [{\citenamefont {Wen}(2004)}]{Wen_book}%
  \BibitemOpen
  \bibfield  {author} {\bibinfo {author} {\bibfnamefont {X.-G.}\ \bibnamefont
  {Wen}},\ }\href@noop {} {\bibfield  {journal} {\bibinfo  {journal} {Quantum
  Field Theory of Many-body Systems from the Origin of Sound to an Origin of
  Light and Electrons by Xiao-Gang Wen. Published in the United States by
  Oxford University Press Inc., New York, 2004. ISBN 019853094.}\ }\textbf
  {\bibinfo {volume} {1}} (\bibinfo {year} {2004})}\BibitemShut {NoStop}%
\bibitem [{\citenamefont {Chen}\ \emph {et~al.}(2010)\citenamefont {Chen},
  \citenamefont {Gu},\ and\ \citenamefont {Wen}}]{ref:Wen-topo}%
  \BibitemOpen
  \bibfield  {author} {\bibinfo {author} {\bibfnamefont {X.}~\bibnamefont
  {Chen}}, \bibinfo {author} {\bibfnamefont {Z.-C.}\ \bibnamefont {Gu}}, \ and\
  \bibinfo {author} {\bibfnamefont {X.-G.}\ \bibnamefont {Wen}},\ }\href
  {\doibase 10.1103/PhysRevB.82.155138} {\bibfield  {journal} {\bibinfo
  {journal} {Phys. Rev. B}\ }\textbf {\bibinfo {volume} {82}},\ \bibinfo
  {pages} {155138} (\bibinfo {year} {2010})}\BibitemShut {NoStop}%
\bibitem [{\citenamefont {Hayden}\ \emph {et~al.}(2006)\citenamefont {Hayden},
  \citenamefont {Leung},\ and\ \citenamefont {Winter}}]{hayden2006aspects}%
  \BibitemOpen
  \bibfield  {author} {\bibinfo {author} {\bibfnamefont {P.}~\bibnamefont
  {Hayden}}, \bibinfo {author} {\bibfnamefont {W.~D.}\ \bibnamefont {Leung}}, \
  and\ \bibinfo {author} {\bibfnamefont {A.}~\bibnamefont {Winter}},\ }\href
  {\doibase 10.1007/s00220-006-1535-6} {\bibfield  {journal} {\bibinfo
  {journal} {Communications in Mathematical Physics}\ }\textbf {\bibinfo
  {volume} {265}},\ \bibinfo {pages} {95} (\bibinfo {year} {2006})}\BibitemShut
  {NoStop}%
\bibitem [{\citenamefont {{Osborne}}(2012)}]{ref:Osborne-QHC12}%
  \BibitemOpen
  \bibfield  {author} {\bibinfo {author} {\bibfnamefont {T.~J.}\ \bibnamefont
  {{Osborne}}},\ }\href {\doibase 10.1088/0034-4885/75/2/022001} {\bibfield
  {journal} {\bibinfo  {journal} {Reports on Progress in Physics}\ }\textbf
  {\bibinfo {volume} {75}},\ \bibinfo {eid} {022001} (\bibinfo {year}
  {2012})},\ \Eprint {http://arxiv.org/abs/arXiv:1106.5875}
  {arXiv:arXiv:1106.5875 [quant-ph]} \BibitemShut {NoStop}%
\bibitem [{\citenamefont {Hastings}(2004)}]{ref:Hastings04-LSM}%
  \BibitemOpen
  \bibfield  {author} {\bibinfo {author} {\bibfnamefont {M.~B.}\ \bibnamefont
  {Hastings}},\ }\href {\doibase 10.1103/PhysRevB.69.104431} {\bibfield
  {journal} {\bibinfo  {journal} {Phys. Rev. B}\ }\textbf {\bibinfo {volume}
  {69}},\ \bibinfo {pages} {104431} (\bibinfo {year} {2004})},\ \Eprint
  {http://arxiv.org/abs/arXiv:cond-mat/0305505} {arXiv:cond-mat/0305505}
  \BibitemShut {NoStop}%
\bibitem [{\citenamefont {Hastings}\ and\ \citenamefont
  {Koma}(2006)}]{ref:Hastings06-Expdec}%
  \BibitemOpen
  \bibfield  {author} {\bibinfo {author} {\bibfnamefont {M.~B.}\ \bibnamefont
  {Hastings}}\ and\ \bibinfo {author} {\bibfnamefont {T.}~\bibnamefont
  {Koma}},\ }\href {\doibase 10.1007/s00220-006-0030-4} {\bibfield  {journal}
  {\bibinfo  {journal} {Communications in Mathematical Physics}\ }\textbf
  {\bibinfo {volume} {265}},\ \bibinfo {pages} {781} (\bibinfo {year}
  {2006})}\BibitemShut {NoStop}%
\bibitem [{\citenamefont {Nachtergaele}\ and\ \citenamefont
  {Sims}(2006)}]{ref:LR-Nachtergaele06}%
  \BibitemOpen
  \bibfield  {author} {\bibinfo {author} {\bibfnamefont {B.}~\bibnamefont
  {Nachtergaele}}\ and\ \bibinfo {author} {\bibfnamefont {R.}~\bibnamefont
  {Sims}},\ }\href {\doibase 10.1007/s00220-006-1556-1} {\bibfield  {journal}
  {\bibinfo  {journal} {Communications in Mathematical Physics}\ }\textbf
  {\bibinfo {volume} {265}},\ \bibinfo {pages} {119} (\bibinfo {year}
  {2006})}\BibitemShut {NoStop}%
\bibitem [{\citenamefont {Eisert}\ \emph {et~al.}(2010)\citenamefont {Eisert},
  \citenamefont {Cramer},\ and\ \citenamefont {Plenio}}]{ref:AL-rev}%
  \BibitemOpen
  \bibfield  {author} {\bibinfo {author} {\bibfnamefont {J.}~\bibnamefont
  {Eisert}}, \bibinfo {author} {\bibfnamefont {M.}~\bibnamefont {Cramer}}, \
  and\ \bibinfo {author} {\bibfnamefont {M.~B.}\ \bibnamefont {Plenio}},\
  }\href {\doibase 10.1103/RevModPhys.82.277} {\bibfield  {journal} {\bibinfo
  {journal} {Rev. Mod. Phys.}\ }\textbf {\bibinfo {volume} {82}},\ \bibinfo
  {pages} {277} (\bibinfo {year} {2010})},\ \Eprint
  {http://arxiv.org/abs/arXiv:0808.3773} {arXiv:0808.3773} \BibitemShut
  {NoStop}%
\bibitem [{\citenamefont {Amico}\ \emph {et~al.}(2008)\citenamefont {Amico},
  \citenamefont {Fazio}, \citenamefont {Osterloh},\ and\ \citenamefont
  {Vedral}}]{ref:Ent-rev-2008}%
  \BibitemOpen
  \bibfield  {author} {\bibinfo {author} {\bibfnamefont {L.}~\bibnamefont
  {Amico}}, \bibinfo {author} {\bibfnamefont {R.}~\bibnamefont {Fazio}},
  \bibinfo {author} {\bibfnamefont {A.}~\bibnamefont {Osterloh}}, \ and\
  \bibinfo {author} {\bibfnamefont {V.}~\bibnamefont {Vedral}},\ }\href
  {\doibase 10.1103/RevModPhys.80.517} {\bibfield  {journal} {\bibinfo
  {journal} {Rev. Mod. Phys.}\ }\textbf {\bibinfo {volume} {80}},\ \bibinfo
  {pages} {517} (\bibinfo {year} {2008})}\BibitemShut {NoStop}%
\bibitem [{\citenamefont {Hastings}(2007{\natexlab{a}})}]{ref:Hastings-AL07}%
  \BibitemOpen
  \bibfield  {author} {\bibinfo {author} {\bibfnamefont {M.~B.}\ \bibnamefont
  {Hastings}},\ }\href {http://stacks.iop.org/1742-5468/2007/i=08/a=P08024}
  {\bibfield  {journal} {\bibinfo  {journal} {Journal of Statistical Mechanics:
  Theory and Experiment}\ }\textbf {\bibinfo {volume} {2007}},\ \bibinfo
  {pages} {P08024} (\bibinfo {year} {2007}{\natexlab{a}})},\ \Eprint
  {http://arxiv.org/abs/arXiv:0705.2024} {arXiv:0705.2024} \BibitemShut
  {NoStop}%
\bibitem [{\citenamefont {Arad}\ \emph {et~al.}(2012)\citenamefont {Arad},
  \citenamefont {Landau},\ and\ \citenamefont {Vazirani}}]{ref:Arad-AL12}%
  \BibitemOpen
  \bibfield  {author} {\bibinfo {author} {\bibfnamefont {I.}~\bibnamefont
  {Arad}}, \bibinfo {author} {\bibfnamefont {Z.}~\bibnamefont {Landau}}, \ and\
  \bibinfo {author} {\bibfnamefont {U.}~\bibnamefont {Vazirani}},\ }\href
  {\doibase 10.1103/PhysRevB.85.195145} {\bibfield  {journal} {\bibinfo
  {journal} {Phys. Rev. B}\ }\textbf {\bibinfo {volume} {85}},\ \bibinfo
  {pages} {195145} (\bibinfo {year} {2012})}\BibitemShut {NoStop}%
\bibitem [{\citenamefont {Arad}\ \emph {et~al.}(2013)\citenamefont {Arad},
  \citenamefont {Kitaev}, \citenamefont {Landau},\ and\ \citenamefont
  {Vazirani}}]{ref:Arad-AL13}%
  \BibitemOpen
  \bibfield  {author} {\bibinfo {author} {\bibfnamefont {I.}~\bibnamefont
  {Arad}}, \bibinfo {author} {\bibfnamefont {A.}~\bibnamefont {Kitaev}},
  \bibinfo {author} {\bibfnamefont {Z.}~\bibnamefont {Landau}}, \ and\ \bibinfo
  {author} {\bibfnamefont {U.}~\bibnamefont {Vazirani}},\ }\href@noop {}
  {\bibfield  {journal} {\bibinfo  {journal} {ArXiv:1301.1162}\ } (\bibinfo
  {year} {2013})},\ \Eprint {http://arxiv.org/abs/arXiv:1301.1162}
  {arXiv:1301.1162} \BibitemShut {NoStop}%
\bibitem [{\citenamefont {Cho}(2014)}]{cho2014sufficient}%
  \BibitemOpen
  \bibfield  {author} {\bibinfo {author} {\bibfnamefont {J.}~\bibnamefont
  {Cho}},\ }\href {\doibase 10.1103/PhysRevLett.113.197204} {\bibfield
  {journal} {\bibinfo  {journal} {Phys. Rev. Lett.}\ }\textbf {\bibinfo
  {volume} {113}},\ \bibinfo {pages} {197204} (\bibinfo {year}
  {2014})}\BibitemShut {NoStop}%
\bibitem [{\citenamefont {Brand{\~a}o}\ and\ \citenamefont
  {Horodecki}(2013)}]{Brandao}%
  \BibitemOpen
  \bibfield  {author} {\bibinfo {author} {\bibfnamefont {F.~G.}\ \bibnamefont
  {Brand{\~a}o}}\ and\ \bibinfo {author} {\bibfnamefont {M.}~\bibnamefont
  {Horodecki}},\ }\href {\doibase 10.1038/nphys2747} {\bibfield  {journal}
  {\bibinfo  {journal} {Nature Physics}\ }\textbf {\bibinfo {volume} {9}},\
  \bibinfo {pages} {721} (\bibinfo {year} {2013})}\BibitemShut {NoStop}%
\bibitem [{\citenamefont {Arad}\ \emph
  {et~al.}(2016{\natexlab{a}})\citenamefont {Arad}, \citenamefont {Landau},
  \citenamefont {Vazirani},\ and\ \citenamefont {Vidick}}]{arad2016rigorous}%
  \BibitemOpen
  \bibfield  {author} {\bibinfo {author} {\bibfnamefont {I.}~\bibnamefont
  {Arad}}, \bibinfo {author} {\bibfnamefont {Z.}~\bibnamefont {Landau}},
  \bibinfo {author} {\bibfnamefont {U.}~\bibnamefont {Vazirani}}, \ and\
  \bibinfo {author} {\bibfnamefont {T.}~\bibnamefont {Vidick}},\ }\href@noop {}
  {\bibfield  {journal} {\bibinfo  {journal} {arXiv preprint arXiv:1602.08828}\
  } (\bibinfo {year} {2016}{\natexlab{a}})},\ \Eprint
  {http://arxiv.org/abs/arXiv:1602.08828} {arXiv:1602.08828} \BibitemShut
  {NoStop}%
\bibitem [{\citenamefont {Verstraete}\ \emph {et~al.}(2008)\citenamefont
  {Verstraete}, \citenamefont {Murg},\ and\ \citenamefont
  {Cirac}}]{verstraete2008matrix}%
  \BibitemOpen
  \bibfield  {author} {\bibinfo {author} {\bibfnamefont {F.}~\bibnamefont
  {Verstraete}}, \bibinfo {author} {\bibfnamefont {V.}~\bibnamefont {Murg}}, \
  and\ \bibinfo {author} {\bibfnamefont {J.~I.}\ \bibnamefont {Cirac}},\
  }\href@noop {} {\bibfield  {journal} {\bibinfo  {journal} {Advances in
  Physics}\ }\textbf {\bibinfo {volume} {57}},\ \bibinfo {pages} {143}
  (\bibinfo {year} {2008})}\BibitemShut {NoStop}%
\bibitem [{\citenamefont {Chen}\ \emph {et~al.}(2011)\citenamefont {Chen},
  \citenamefont {Gu},\ and\ \citenamefont {Wen}}]{ref:Wen10}%
  \BibitemOpen
  \bibfield  {author} {\bibinfo {author} {\bibfnamefont {X.}~\bibnamefont
  {Chen}}, \bibinfo {author} {\bibfnamefont {Z.-C.}\ \bibnamefont {Gu}}, \ and\
  \bibinfo {author} {\bibfnamefont {X.-G.}\ \bibnamefont {Wen}},\ }\href
  {\doibase 10.1103/PhysRevB.84.235128} {\bibfield  {journal} {\bibinfo
  {journal} {Phys. Rev. B}\ }\textbf {\bibinfo {volume} {84}},\ \bibinfo
  {pages} {235128} (\bibinfo {year} {2011})}\BibitemShut {NoStop}%
\bibitem [{\citenamefont {Ge}\ and\ \citenamefont
  {Eisert}(2016)}]{ref:Ge2014-AL-no-PEPS}%
  \BibitemOpen
  \bibfield  {author} {\bibinfo {author} {\bibfnamefont {Y.}~\bibnamefont
  {Ge}}\ and\ \bibinfo {author} {\bibfnamefont {J.}~\bibnamefont {Eisert}},\
  }\href {http://stacks.iop.org/1367-2630/18/i=8/a=083026} {\bibfield
  {journal} {\bibinfo  {journal} {New Journal of Physics}\ }\textbf {\bibinfo
  {volume} {18}},\ \bibinfo {pages} {083026} (\bibinfo {year} {2016})},\
  \Eprint {http://arxiv.org/abs/arXiv:1411.2995} {arXiv:1411.2995} \BibitemShut
  {NoStop}%
\bibitem [{\citenamefont {Hoory}\ \emph {et~al.}(2006)\citenamefont {Hoory},
  \citenamefont {Linial},\ and\ \citenamefont {Wigderson}}]{hoory2006expander}%
  \BibitemOpen
  \bibfield  {author} {\bibinfo {author} {\bibfnamefont {S.}~\bibnamefont
  {Hoory}}, \bibinfo {author} {\bibfnamefont {N.}~\bibnamefont {Linial}}, \
  and\ \bibinfo {author} {\bibfnamefont {A.}~\bibnamefont {Wigderson}},\ }\href
  {\doibase 10.1090/S0273-0979-06-01126-8} {\bibfield  {journal} {\bibinfo
  {journal} {Bulletin of the American Mathematical Society}\ }\textbf {\bibinfo
  {volume} {43}},\ \bibinfo {pages} {439} (\bibinfo {year} {2006})}\BibitemShut
  {NoStop}%
\bibitem [{\citenamefont {Kempe}\ \emph {et~al.}(2006)\citenamefont {Kempe},
  \citenamefont {Kitaev},\ and\ \citenamefont {Regev}}]{ref:KempeKitaevRegev}%
  \BibitemOpen
  \bibfield  {author} {\bibinfo {author} {\bibfnamefont {J.}~\bibnamefont
  {Kempe}}, \bibinfo {author} {\bibfnamefont {A.}~\bibnamefont {Kitaev}}, \
  and\ \bibinfo {author} {\bibfnamefont {O.}~\bibnamefont {Regev}},\
  }\href@noop {} {\bibfield  {journal} {\bibinfo  {journal} {SIAM Journal on
  Computing}\ }\textbf {\bibinfo {volume} {35}},\ \bibinfo {pages} {1070}
  (\bibinfo {year} {2006})},\ \Eprint
  {http://arxiv.org/abs/arXiv:quant-ph/0406180} {arXiv:quant-ph/0406180}
  \BibitemShut {NoStop}%
\bibitem [{\citenamefont {Hastings}(2007{\natexlab{b}})}]{PhysRevB.76.035114}%
  \BibitemOpen
  \bibfield  {author} {\bibinfo {author} {\bibfnamefont {M.~B.}\ \bibnamefont
  {Hastings}},\ }\href {\doibase 10.1103/PhysRevB.76.035114} {\bibfield
  {journal} {\bibinfo  {journal} {Phys. Rev. B}\ }\textbf {\bibinfo {volume}
  {76}},\ \bibinfo {pages} {035114} (\bibinfo {year}
  {2007}{\natexlab{b}})}\BibitemShut {NoStop}%
\bibitem [{\citenamefont {Aharonov}\ \emph {et~al.}(2013)\citenamefont
  {Aharonov}, \citenamefont {Arad},\ and\ \citenamefont
  {Vidick}}]{aharonov2013guest}%
  \BibitemOpen
  \bibfield  {author} {\bibinfo {author} {\bibfnamefont {D.}~\bibnamefont
  {Aharonov}}, \bibinfo {author} {\bibfnamefont {I.}~\bibnamefont {Arad}}, \
  and\ \bibinfo {author} {\bibfnamefont {T.}~\bibnamefont {Vidick}},\
  }\href@noop {} {\bibfield  {journal} {\bibinfo  {journal} {Acm sigact news}\
  }\textbf {\bibinfo {volume} {44}},\ \bibinfo {pages} {47} (\bibinfo {year}
  {2013})},\ \Eprint {http://arxiv.org/abs/arXiv:1309.7495} {arXiv:1309.7495}
  \BibitemShut {NoStop}%
\bibitem [{\citenamefont {Chen}\ \emph {et~al.}(2013)\citenamefont {Chen},
  \citenamefont {Gu}, \citenamefont {Liu},\ and\ \citenamefont
  {Wen}}]{chen2011symmetry}%
  \BibitemOpen
  \bibfield  {author} {\bibinfo {author} {\bibfnamefont {X.}~\bibnamefont
  {Chen}}, \bibinfo {author} {\bibfnamefont {Z.-C.}\ \bibnamefont {Gu}},
  \bibinfo {author} {\bibfnamefont {Z.-X.}\ \bibnamefont {Liu}}, \ and\
  \bibinfo {author} {\bibfnamefont {X.-G.}\ \bibnamefont {Wen}},\ }\href
  {\doibase 10.1103/PhysRevB.87.155114} {\bibfield  {journal} {\bibinfo
  {journal} {Phys. Rev. B}\ }\textbf {\bibinfo {volume} {87}},\ \bibinfo
  {pages} {155114} (\bibinfo {year} {2013})}\BibitemShut {NoStop}%
\bibitem [{\citenamefont {Shimizu}\ and\ \citenamefont
  {Morimae}(2005)}]{ref:Shimizu05-macroE}%
  \BibitemOpen
  \bibfield  {author} {\bibinfo {author} {\bibfnamefont {A.}~\bibnamefont
  {Shimizu}}\ and\ \bibinfo {author} {\bibfnamefont {T.}~\bibnamefont
  {Morimae}},\ }\href {\doibase 10.1103/PhysRevLett.95.090401} {\bibfield
  {journal} {\bibinfo  {journal} {Phys. Rev. Lett.}\ }\textbf {\bibinfo
  {volume} {95}},\ \bibinfo {pages} {090401} (\bibinfo {year}
  {2005})}\BibitemShut {NoStop}%
\bibitem [{\citenamefont {Fr{\"o}wis}\ and\ \citenamefont
  {D{\"u}r}(2012)}]{frowis2012}%
  \BibitemOpen
  \bibfield  {author} {\bibinfo {author} {\bibfnamefont {F.}~\bibnamefont
  {Fr{\"o}wis}}\ and\ \bibinfo {author} {\bibfnamefont {W.}~\bibnamefont
  {D{\"u}r}},\ }\href {http://stacks.iop.org/1367-2630/14/i=9/a=093039}
  {\bibfield  {journal} {\bibinfo  {journal} {New Journal of Physics}\ }\textbf
  {\bibinfo {volume} {14}},\ \bibinfo {pages} {093039} (\bibinfo {year}
  {2012})}\BibitemShut {NoStop}%
\bibitem [{\citenamefont {Haldane}(1983)}]{ref:Heisen}%
  \BibitemOpen
  \bibfield  {author} {\bibinfo {author} {\bibfnamefont {F.~D.~M.}\
  \bibnamefont {Haldane}},\ }\href {\doibase 10.1103/PhysRevLett.50.1153}
  {\bibfield  {journal} {\bibinfo  {journal} {Phys. Rev. Lett.}\ }\textbf
  {\bibinfo {volume} {50}},\ \bibinfo {pages} {1153} (\bibinfo {year}
  {1983})}\BibitemShut {NoStop}%
\bibitem [{\citenamefont {Affleck}\ \emph {et~al.}(1987)\citenamefont
  {Affleck}, \citenamefont {Kennedy}, \citenamefont {Lieb},\ and\ \citenamefont
  {Tasaki}}]{ref:AKLT}%
  \BibitemOpen
  \bibfield  {author} {\bibinfo {author} {\bibfnamefont {I.}~\bibnamefont
  {Affleck}}, \bibinfo {author} {\bibfnamefont {T.}~\bibnamefont {Kennedy}},
  \bibinfo {author} {\bibfnamefont {E.~H.}\ \bibnamefont {Lieb}}, \ and\
  \bibinfo {author} {\bibfnamefont {H.}~\bibnamefont {Tasaki}},\ }\href
  {\doibase 10.1103/PhysRevLett.59.799} {\bibfield  {journal} {\bibinfo
  {journal} {Phys. Rev. Lett.}\ }\textbf {\bibinfo {volume} {59}},\ \bibinfo
  {pages} {799} (\bibinfo {year} {1987})}\BibitemShut {NoStop}%
\bibitem [{\citenamefont {Lipkin}\ \emph {et~al.}(1965)\citenamefont {Lipkin},
  \citenamefont {Meshkov},\ and\ \citenamefont {Glick}}]{LMG1}%
  \BibitemOpen
  \bibfield  {author} {\bibinfo {author} {\bibfnamefont {H.}~\bibnamefont
  {Lipkin}}, \bibinfo {author} {\bibfnamefont {N.}~\bibnamefont {Meshkov}}, \
  and\ \bibinfo {author} {\bibfnamefont {A.}~\bibnamefont {Glick}},\ }\href
  {\doibase 10.1016/0029-5582(65)90862-X} {\bibfield  {journal} {\bibinfo
  {journal} {Nuclear Physics}\ }\textbf {\bibinfo {volume} {62}},\ \bibinfo
  {pages} {188 } (\bibinfo {year} {1965})}\BibitemShut {NoStop}%
\bibitem [{\citenamefont {Cui}\ \emph {et~al.}(2013)\citenamefont {Cui},
  \citenamefont {Amico}, \citenamefont {Fan}, \citenamefont {Gu}, \citenamefont
  {Hamma},\ and\ \citenamefont {Vedral}}]{ref:Cluster_ising}%
  \BibitemOpen
  \bibfield  {author} {\bibinfo {author} {\bibfnamefont {J.}~\bibnamefont
  {Cui}}, \bibinfo {author} {\bibfnamefont {L.}~\bibnamefont {Amico}}, \bibinfo
  {author} {\bibfnamefont {H.}~\bibnamefont {Fan}}, \bibinfo {author}
  {\bibfnamefont {M.}~\bibnamefont {Gu}}, \bibinfo {author} {\bibfnamefont
  {A.}~\bibnamefont {Hamma}}, \ and\ \bibinfo {author} {\bibfnamefont
  {V.}~\bibnamefont {Vedral}},\ }\href {\doibase 10.1103/PhysRevB.88.125117}
  {\bibfield  {journal} {\bibinfo  {journal} {Phys. Rev. B}\ }\textbf {\bibinfo
  {volume} {88}},\ \bibinfo {pages} {125117} (\bibinfo {year}
  {2013})}\BibitemShut {NoStop}%
\bibitem [{\citenamefont {Kitaev}(2003)}]{ref:Kitaev03-toric}%
  \BibitemOpen
  \bibfield  {author} {\bibinfo {author} {\bibfnamefont {A.}~\bibnamefont
  {Kitaev}},\ }\href {\doibase http://dx.doi.org/10.1016/S0003-4916(02)00018-0}
  {\bibfield  {journal} {\bibinfo  {journal} {Annals of Physics}\ }\textbf
  {\bibinfo {volume} {303}},\ \bibinfo {pages} {2 } (\bibinfo {year}
  {2003})}\BibitemShut {NoStop}%
\bibitem [{\citenamefont {Levin}\ and\ \citenamefont
  {Wen}(2005)}]{ref:Levin-Wen}%
  \BibitemOpen
  \bibfield  {author} {\bibinfo {author} {\bibfnamefont {M.~A.}\ \bibnamefont
  {Levin}}\ and\ \bibinfo {author} {\bibfnamefont {X.-G.}\ \bibnamefont
  {Wen}},\ }\href {\doibase 10.1103/PhysRevB.71.045110} {\bibfield  {journal}
  {\bibinfo  {journal} {Phys. Rev. B}\ }\textbf {\bibinfo {volume} {71}},\
  \bibinfo {pages} {045110} (\bibinfo {year} {2005})}\BibitemShut {NoStop}%
\bibitem [{\citenamefont {Arad}\ \emph
  {et~al.}(2016{\natexlab{b}})\citenamefont {Arad}, \citenamefont {Kuwahara},\
  and\ \citenamefont {Landau}}]{ref:Arad14-Edist}%
  \BibitemOpen
  \bibfield  {author} {\bibinfo {author} {\bibfnamefont {I.}~\bibnamefont
  {Arad}}, \bibinfo {author} {\bibfnamefont {T.}~\bibnamefont {Kuwahara}}, \
  and\ \bibinfo {author} {\bibfnamefont {Z.}~\bibnamefont {Landau}},\ }\href
  {http://stacks.iop.org/1742-5468/2016/i=3/a=033301} {\bibfield  {journal}
  {\bibinfo  {journal} {Journal of Statistical Mechanics: Theory and
  Experiment}\ }\textbf {\bibinfo {volume} {2016}},\ \bibinfo {pages} {033301}
  (\bibinfo {year} {2016}{\natexlab{b}})},\ \Eprint
  {http://arxiv.org/abs/arXiv:1406.3898} {arXiv:1406.3898} \BibitemShut
  {NoStop}%
\bibitem [{\citenamefont {Raussendorf}\ and\ \citenamefont
  {Briegel}(2001)}]{raussendorf2001one}%
  \BibitemOpen
  \bibfield  {author} {\bibinfo {author} {\bibfnamefont {R.}~\bibnamefont
  {Raussendorf}}\ and\ \bibinfo {author} {\bibfnamefont {H.~J.}\ \bibnamefont
  {Briegel}},\ }\href {\doibase 10.1103/PhysRevLett.86.5188} {\bibfield
  {journal} {\bibinfo  {journal} {Phys. Rev. Lett.}\ }\textbf {\bibinfo
  {volume} {86}},\ \bibinfo {pages} {5188} (\bibinfo {year}
  {2001})}\BibitemShut {NoStop}%
\bibitem [{\citenamefont {Briegel}\ and\ \citenamefont
  {Raussendorf}(2001)}]{briegel2001persistent}%
  \BibitemOpen
  \bibfield  {author} {\bibinfo {author} {\bibfnamefont {H.~J.}\ \bibnamefont
  {Briegel}}\ and\ \bibinfo {author} {\bibfnamefont {R.}~\bibnamefont
  {Raussendorf}},\ }\href {\doibase 10.1103/PhysRevLett.86.910} {\bibfield
  {journal} {\bibinfo  {journal} {Phys. Rev. Lett.}\ }\textbf {\bibinfo
  {volume} {86}},\ \bibinfo {pages} {910} (\bibinfo {year} {2001})}\BibitemShut
  {NoStop}%
\bibitem [{\citenamefont {Hein}\ \emph {et~al.}(2004)\citenamefont {Hein},
  \citenamefont {Eisert},\ and\ \citenamefont {Briegel}}]{hein2004multiparty}%
  \BibitemOpen
  \bibfield  {author} {\bibinfo {author} {\bibfnamefont {M.}~\bibnamefont
  {Hein}}, \bibinfo {author} {\bibfnamefont {J.}~\bibnamefont {Eisert}}, \ and\
  \bibinfo {author} {\bibfnamefont {H.~J.}\ \bibnamefont {Briegel}},\ }\href
  {\doibase 10.1103/PhysRevA.69.062311} {\bibfield  {journal} {\bibinfo
  {journal} {Phys. Rev. A}\ }\textbf {\bibinfo {volume} {69}},\ \bibinfo
  {pages} {062311} (\bibinfo {year} {2004})}\BibitemShut {NoStop}%
\bibitem [{\citenamefont {Hastings}(2011)}]{ref:Hastings10-nonZeroT}%
  \BibitemOpen
  \bibfield  {author} {\bibinfo {author} {\bibfnamefont {M.~B.}\ \bibnamefont
  {Hastings}},\ }\href {\doibase 10.1103/PhysRevLett.107.210501} {\bibfield
  {journal} {\bibinfo  {journal} {Phys. Rev. Lett.}\ }\textbf {\bibinfo
  {volume} {107}},\ \bibinfo {pages} {210501} (\bibinfo {year} {2011})},\
  \Eprint {http://arxiv.org/abs/arXiv:1106.6026} {arXiv:1106.6026} \BibitemShut
  {NoStop}%
\bibitem [{\citenamefont {Freedman}\ and\ \citenamefont
  {Hastings}(2014)}]{ref:Hastings14-NLTS}%
  \BibitemOpen
  \bibfield  {author} {\bibinfo {author} {\bibfnamefont {M.~H.}\ \bibnamefont
  {Freedman}}\ and\ \bibinfo {author} {\bibfnamefont {M.~B.}\ \bibnamefont
  {Hastings}},\ }\href@noop {} {\bibfield  {journal} {\bibinfo  {journal}
  {Quantum Information \& Computation}\ }\textbf {\bibinfo {volume} {14}},\
  \bibinfo {pages} {144} (\bibinfo {year} {2014})}\BibitemShut {NoStop}%
\bibitem [{\citenamefont {Hastings}(2012)}]{ref:Toric-on-a-sphere}%
  \BibitemOpen
  \bibfield  {author} {\bibinfo {author} {\bibfnamefont {M.~B.}\ \bibnamefont
  {Hastings}},\ }\enquote {\bibinfo {title} {Locality in quantum systems},}\
  in\ \href@noop {} {\emph {\bibinfo {booktitle} {Quantum Theory from Small to
  Large Scales: Lecture Notes of the Les Houches Summer School: Volume 95,
  August 2010}}}\ (\bibinfo  {publisher} {Oxford Scholarship Online},\ \bibinfo
  {year} {2012})\ p.\ \bibinfo {pages} {see pg.~24 on arXiv version
  arXiv:1008.5137},\ \Eprint {http://arxiv.org/abs/arXiv:1008.5137}
  {arXiv:1008.5137} \BibitemShut {NoStop}%
\bibitem [{\citenamefont {Zhou}(2008)}]{irred:Zhou}%
  \BibitemOpen
  \bibfield  {author} {\bibinfo {author} {\bibfnamefont {D.~L.}\ \bibnamefont
  {Zhou}},\ }\href {\doibase 10.1103/PhysRevLett.101.180505} {\bibfield
  {journal} {\bibinfo  {journal} {Phys. Rev. Lett.}\ }\textbf {\bibinfo
  {volume} {101}},\ \bibinfo {pages} {180505} (\bibinfo {year}
  {2008})}\BibitemShut {NoStop}%
\bibitem [{\citenamefont {Kato}\ \emph {et~al.}(2016)\citenamefont {Kato},
  \citenamefont {Furrer},\ and\ \citenamefont {Murao}}]{Kato:cor}%
  \BibitemOpen
  \bibfield  {author} {\bibinfo {author} {\bibfnamefont {K.}~\bibnamefont
  {Kato}}, \bibinfo {author} {\bibfnamefont {F.}~\bibnamefont {Furrer}}, \ and\
  \bibinfo {author} {\bibfnamefont {M.}~\bibnamefont {Murao}},\ }\href
  {\doibase 10.1103/PhysRevA.93.022317} {\bibfield  {journal} {\bibinfo
  {journal} {Phys. Rev. A}\ }\textbf {\bibinfo {volume} {93}},\ \bibinfo
  {pages} {022317} (\bibinfo {year} {2016})}\BibitemShut {NoStop}%
\bibitem [{\citenamefont {Liu}\ \emph {et~al.}(2016)\citenamefont {Liu},
  \citenamefont {Zeng},\ and\ \citenamefont {Zhou}}]{liu2014irreducible}%
  \BibitemOpen
  \bibfield  {author} {\bibinfo {author} {\bibfnamefont {Y.}~\bibnamefont
  {Liu}}, \bibinfo {author} {\bibfnamefont {B.}~\bibnamefont {Zeng}}, \ and\
  \bibinfo {author} {\bibfnamefont {D.~L.}\ \bibnamefont {Zhou}},\ }\href
  {http://stacks.iop.org/1367-2630/18/i=2/a=023024} {\bibfield  {journal}
  {\bibinfo  {journal} {New Journal of Physics}\ }\textbf {\bibinfo {volume}
  {18}},\ \bibinfo {pages} {023024} (\bibinfo {year} {2016})},\ \Eprint
  {http://arxiv.org/abs/arXiv:1402.4245} {arXiv:1402.4245} \BibitemShut
  {NoStop}%
\bibitem [{\citenamefont {Son}\ \emph {et~al.}(2012)\citenamefont {Son},
  \citenamefont {Amico},\ and\ \citenamefont {Vedral}}]{son2012topological}%
  \BibitemOpen
  \bibfield  {author} {\bibinfo {author} {\bibfnamefont {W.}~\bibnamefont
  {Son}}, \bibinfo {author} {\bibfnamefont {L.}~\bibnamefont {Amico}}, \ and\
  \bibinfo {author} {\bibfnamefont {V.}~\bibnamefont {Vedral}},\ }\href@noop {}
  {\bibfield  {journal} {\bibinfo  {journal} {Quantum Information Processing}\
  }\textbf {\bibinfo {volume} {11}},\ \bibinfo {pages} {1961} (\bibinfo {year}
  {2012})}\BibitemShut {NoStop}%
\bibitem [{\citenamefont {Hoeffding}(1963)}]{ref:Hoeffding-ineq}%
  \BibitemOpen
  \bibfield  {author} {\bibinfo {author} {\bibfnamefont {W.}~\bibnamefont
  {Hoeffding}},\ }\href {\doibase 10.1080/01621459.1963.10500830} {\bibfield
  {journal} {\bibinfo  {journal} {Journal of the American Statistical
  Association}\ }\textbf {\bibinfo {volume} {58}},\ \bibinfo {pages} {13}
  (\bibinfo {year} {1963})}\BibitemShut {NoStop}%
\bibitem [{\citenamefont {Kuwahara}(2016)}]{JSTAT_extend}%
  \BibitemOpen
  \bibfield  {author} {\bibinfo {author} {\bibfnamefont {T.}~\bibnamefont
  {Kuwahara}},\ }\href {http://stacks.iop.org/1742-5468/2016/i=5/a=053103}
  {\bibfield  {journal} {\bibinfo  {journal} {Journal of Statistical Mechanics:
  Theory and Experiment}\ }\textbf {\bibinfo {volume} {2016}},\ \bibinfo
  {pages} {053103} (\bibinfo {year} {2016})}\BibitemShut {NoStop}%
\bibitem [{\citenamefont {Continentino}(1994)}]{ref:Conti94-Quantum-scaling}%
  \BibitemOpen
  \bibfield  {author} {\bibinfo {author} {\bibfnamefont {M.~A.}\ \bibnamefont
  {Continentino}},\ }\href {\doibase
  http://dx.doi.org/10.1016/0370-1573(94)90112-0} {\bibfield  {journal}
  {\bibinfo  {journal} {Physics Reports}\ }\textbf {\bibinfo {volume} {239}},\
  \bibinfo {pages} {179 } (\bibinfo {year} {1994})}\BibitemShut {NoStop}%
\bibitem [{\citenamefont {Vojta}\ and\ \citenamefont
  {Schmalian}(2005)}]{vojta2005percolation}%
  \BibitemOpen
  \bibfield  {author} {\bibinfo {author} {\bibfnamefont {T.}~\bibnamefont
  {Vojta}}\ and\ \bibinfo {author} {\bibfnamefont {J.}~\bibnamefont
  {Schmalian}},\ }\href {\doibase 10.1103/PhysRevLett.95.237206} {\bibfield
  {journal} {\bibinfo  {journal} {Phys. Rev. Lett.}\ }\textbf {\bibinfo
  {volume} {95}},\ \bibinfo {pages} {237206} (\bibinfo {year}
  {2005})}\BibitemShut {NoStop}%
\bibitem [{\citenamefont {Dusuel}\ and\ \citenamefont
  {Vidal}(2004)}]{dusuel2004finite}%
  \BibitemOpen
  \bibfield  {author} {\bibinfo {author} {\bibfnamefont {S.}~\bibnamefont
  {Dusuel}}\ and\ \bibinfo {author} {\bibfnamefont {J.}~\bibnamefont {Vidal}},\
  }\href {\doibase 10.1103/PhysRevLett.93.237204} {\bibfield  {journal}
  {\bibinfo  {journal} {Phys. Rev. Lett.}\ }\textbf {\bibinfo {volume} {93}},\
  \bibinfo {pages} {237204} (\bibinfo {year} {2004})}\BibitemShut {NoStop}%
\bibitem [{\citenamefont {Brand{\~a}o}\ and\ \citenamefont
  {Harrow}(2013)}]{ref:Brandao13-qpcp}%
  \BibitemOpen
  \bibfield  {author} {\bibinfo {author} {\bibfnamefont {F.~G.}\ \bibnamefont
  {Brand{\~a}o}}\ and\ \bibinfo {author} {\bibfnamefont {A.~W.}\ \bibnamefont
  {Harrow}},\ }in\ \href {\doibase 10.1145/2488608.2488719} {\emph {\bibinfo
  {booktitle} {Proceedings of the Forty-fifth Annual ACM Symposium on Theory of
  Computing}}},\ \bibinfo {series and number} {STOC '13}\ (\bibinfo
  {publisher} {ACM},\ \bibinfo {address} {New York, NY, USA},\ \bibinfo {year}
  {2013})\ pp.\ \bibinfo {pages} {871--880}\BibitemShut {NoStop}%
\bibitem [{\citenamefont {Osterloh}\ and\ \citenamefont
  {Sch\"utzhold}(2015)}]{ref:Osterloh14-MF}%
  \BibitemOpen
  \bibfield  {author} {\bibinfo {author} {\bibfnamefont {A.}~\bibnamefont
  {Osterloh}}\ and\ \bibinfo {author} {\bibfnamefont {R.}~\bibnamefont
  {Sch\"utzhold}},\ }\href {\doibase 10.1103/PhysRevB.91.125114} {\bibfield
  {journal} {\bibinfo  {journal} {Phys. Rev. B}\ }\textbf {\bibinfo {volume}
  {91}},\ \bibinfo {pages} {125114} (\bibinfo {year} {2015})}\BibitemShut
  {NoStop}%
\bibitem [{\citenamefont {{Brand{\~a}o}}\ and\ \citenamefont
  {{Cramer}}(2015)}]{ref:Brandao2015-Berry-Esseen}%
  \BibitemOpen
  \bibfield  {author} {\bibinfo {author} {\bibfnamefont {F.~G.~S.~L.}\
  \bibnamefont {{Brand{\~a}o}}}\ and\ \bibinfo {author} {\bibfnamefont
  {M.}~\bibnamefont {{Cramer}}},\ }\href@noop {} {\bibfield  {journal}
  {\bibinfo  {journal} {ArXiv:1502.03263}\ } (\bibinfo {year} {2015})},\
  \Eprint {http://arxiv.org/abs/arXiv:1502.03263} {arXiv:arXiv:1502.03263
  [quant-ph]} \BibitemShut {NoStop}%
\bibitem [{\citenamefont {Lubasch}\ \emph
  {et~al.}(2014{\natexlab{a}})\citenamefont {Lubasch}, \citenamefont {Cirac},\
  and\ \citenamefont {Ba{\~n}uls}}]{ref:Cirac14-PEPS-alg}%
  \BibitemOpen
  \bibfield  {author} {\bibinfo {author} {\bibfnamefont {M.}~\bibnamefont
  {Lubasch}}, \bibinfo {author} {\bibfnamefont {J.~I.}\ \bibnamefont {Cirac}},
  \ and\ \bibinfo {author} {\bibfnamefont {M.-C.}\ \bibnamefont {Ba{\~n}uls}},\
  }\href {\doibase 10.1103/PhysRevB.90.064425} {\bibfield  {journal} {\bibinfo
  {journal} {Phys. Rev. B}\ }\textbf {\bibinfo {volume} {90}},\ \bibinfo
  {pages} {064425} (\bibinfo {year} {2014}{\natexlab{a}})},\ \Eprint
  {http://arxiv.org/abs/arXiv:1405.3259} {arXiv:1405.3259} \BibitemShut
  {NoStop}%
\bibitem [{\citenamefont {Lubasch}\ \emph
  {et~al.}(2014{\natexlab{b}})\citenamefont {Lubasch}, \citenamefont {Cirac},\
  and\ \citenamefont {Ba{\~n}uls}}]{ref:Cirac13-PEPS-contraction}%
  \BibitemOpen
  \bibfield  {author} {\bibinfo {author} {\bibfnamefont {M.}~\bibnamefont
  {Lubasch}}, \bibinfo {author} {\bibfnamefont {J.~I.}\ \bibnamefont {Cirac}},
  \ and\ \bibinfo {author} {\bibfnamefont {M.-C.}\ \bibnamefont {Ba{\~n}uls}},\
  }\href {http://stacks.iop.org/1367-2630/16/i=3/a=033014} {\bibfield
  {journal} {\bibinfo  {journal} {New Journal of Physics}\ }\textbf {\bibinfo
  {volume} {16}},\ \bibinfo {pages} {033014} (\bibinfo {year}
  {2014}{\natexlab{b}})},\ \Eprint {http://arxiv.org/abs/arXiv:1311.6696}
  {arXiv:1311.6696} \BibitemShut {NoStop}%
\bibitem [{\citenamefont {Anshu}\ \emph {et~al.}(2016)\citenamefont {Anshu},
  \citenamefont {Arad},\ and\ \citenamefont {Jain}}]{anshu2016local}%
  \BibitemOpen
  \bibfield  {author} {\bibinfo {author} {\bibfnamefont {A.}~\bibnamefont
  {Anshu}}, \bibinfo {author} {\bibfnamefont {I.}~\bibnamefont {Arad}}, \ and\
  \bibinfo {author} {\bibfnamefont {A.}~\bibnamefont {Jain}},\ }\href@noop {}
  {\bibfield  {journal} {\bibinfo  {journal} {arXiv preprint arXiv:1603.06049}\
  } (\bibinfo {year} {2016})},\ \Eprint {http://arxiv.org/abs/arXiv:1603.06049}
  {arXiv:1603.06049} \BibitemShut {NoStop}%
\bibitem [{\citenamefont {Schwarz}\ \emph {et~al.}(2016)\citenamefont
  {Schwarz}, \citenamefont {Buerschaper},\ and\ \citenamefont
  {Eisert}}]{schwarz2016approximating}%
  \BibitemOpen
  \bibfield  {author} {\bibinfo {author} {\bibfnamefont {M.}~\bibnamefont
  {Schwarz}}, \bibinfo {author} {\bibfnamefont {O.}~\bibnamefont
  {Buerschaper}}, \ and\ \bibinfo {author} {\bibfnamefont {J.}~\bibnamefont
  {Eisert}},\ }\href@noop {} {\bibfield  {journal} {\bibinfo  {journal} {arXiv
  preprint arXiv:1606.06301}\ } (\bibinfo {year} {2016})},\ \Eprint
  {http://arxiv.org/abs/arXiv:1606.06301} {arXiv:1606.06301} \BibitemShut
  {NoStop}%
\bibitem [{\citenamefont {Facchi}\ \emph {et~al.}(2011)\citenamefont {Facchi},
  \citenamefont {Florio}, \citenamefont {Pascazio},\ and\ \citenamefont
  {Pepe}}]{ref:Facchi11-GHZ}%
  \BibitemOpen
  \bibfield  {author} {\bibinfo {author} {\bibfnamefont {P.}~\bibnamefont
  {Facchi}}, \bibinfo {author} {\bibfnamefont {G.}~\bibnamefont {Florio}},
  \bibinfo {author} {\bibfnamefont {S.}~\bibnamefont {Pascazio}}, \ and\
  \bibinfo {author} {\bibfnamefont {F.~V.}\ \bibnamefont {Pepe}},\ }\href
  {\doibase 10.1103/PhysRevLett.107.260502} {\bibfield  {journal} {\bibinfo
  {journal} {Phys. Rev. Lett.}\ }\textbf {\bibinfo {volume} {107}},\ \bibinfo
  {pages} {260502} (\bibinfo {year} {2011})}\BibitemShut {NoStop}%
\bibitem [{\citenamefont {Abramowitz}\ and\ \citenamefont
  {Stegun}(1972)}]{ref:abramowitz1972handbook}%
  \BibitemOpen
  \bibfield  {author} {\bibinfo {author} {\bibfnamefont {M.}~\bibnamefont
  {Abramowitz}}\ and\ \bibinfo {author} {\bibfnamefont {I.~A.}\ \bibnamefont
  {Stegun}},\ }\href@noop {} {\emph {\bibinfo {title} {Handbook of mathematical
  functions: with formulas, graphs, and mathematical tables}}},\ \bibinfo
  {number} {55}\ (\bibinfo  {publisher} {Courier Dover Publications},\ \bibinfo
  {year} {1972})\BibitemShut {NoStop}%
\bibitem [{\citenamefont {Kitaev}\ and\ \citenamefont
  {Preskill}(2006)}]{topo_ent_kit}%
  \BibitemOpen
  \bibfield  {author} {\bibinfo {author} {\bibfnamefont {A.}~\bibnamefont
  {Kitaev}}\ and\ \bibinfo {author} {\bibfnamefont {J.}~\bibnamefont
  {Preskill}},\ }\href {\doibase 10.1103/PhysRevLett.96.110404} {\bibfield
  {journal} {\bibinfo  {journal} {Phys. Rev. Lett.}\ }\textbf {\bibinfo
  {volume} {96}},\ \bibinfo {pages} {110404} (\bibinfo {year}
  {2006})}\BibitemShut {NoStop}%
\bibitem [{\citenamefont {Levin}\ and\ \citenamefont
  {Wen}(2006)}]{topo_ent_wen}%
  \BibitemOpen
  \bibfield  {author} {\bibinfo {author} {\bibfnamefont {M.}~\bibnamefont
  {Levin}}\ and\ \bibinfo {author} {\bibfnamefont {X.-G.}\ \bibnamefont
  {Wen}},\ }\href {\doibase 10.1103/PhysRevLett.96.110405} {\bibfield
  {journal} {\bibinfo  {journal} {Phys. Rev. Lett.}\ }\textbf {\bibinfo
  {volume} {96}},\ \bibinfo {pages} {110405} (\bibinfo {year}
  {2006})}\BibitemShut {NoStop}%
\bibitem [{\citenamefont {Linden}\ \emph {et~al.}(2002)\citenamefont {Linden},
  \citenamefont {Popescu},\ and\ \citenamefont {Wootters}}]{Linden_irred}%
  \BibitemOpen
  \bibfield  {author} {\bibinfo {author} {\bibfnamefont {N.}~\bibnamefont
  {Linden}}, \bibinfo {author} {\bibfnamefont {S.}~\bibnamefont {Popescu}}, \
  and\ \bibinfo {author} {\bibfnamefont {W.~K.}\ \bibnamefont {Wootters}},\
  }\href {\doibase 10.1103/PhysRevLett.89.207901} {\bibfield  {journal}
  {\bibinfo  {journal} {Phys. Rev. Lett.}\ }\textbf {\bibinfo {volume} {89}},\
  \bibinfo {pages} {207901} (\bibinfo {year} {2002})}\BibitemShut {NoStop}%
\bibitem [{\citenamefont {Bravyi}()}]{Bavyi_unpublished}%
  \BibitemOpen
  \bibfield  {author} {\bibinfo {author} {\bibfnamefont {S.}~\bibnamefont
  {Bravyi}},\ }\href@noop {} {\bibinfo  {journal} {unpublished}\ }\BibitemShut
  {NoStop}%
\bibitem [{\citenamefont {Haah}()}]{Haar_slide}%
  \BibitemOpen
\bibfield  {journal} {  }\bibfield  {author} {\bibinfo {author} {\bibfnamefont
  {J.}~\bibnamefont {Haah}},\ }\href
  {http://www.physics.usyd.edu.au/quantum/Coogee2015/Presentations/Haah.pdf#search='Manybody+entanglement+witness+Coogee%2C+Australia'}
  {\bibinfo  {journal} {Coogee'15, talk slide, 16page}\ }\BibitemShut {NoStop}%
\bibitem [{\citenamefont {Galla}\ and\ \citenamefont
  {G\"uhne}(2012)}]{PhysRevE.85.046209}%
  \BibitemOpen
\bibfield  {journal} {  }\bibfield  {author} {\bibinfo {author} {\bibfnamefont
  {T.}~\bibnamefont {Galla}}\ and\ \bibinfo {author} {\bibfnamefont
  {O.}~\bibnamefont {G\"uhne}},\ }\href {\doibase 10.1103/PhysRevE.85.046209}
  {\bibfield  {journal} {\bibinfo  {journal} {Phys. Rev. E}\ }\textbf {\bibinfo
  {volume} {85}},\ \bibinfo {pages} {046209} (\bibinfo {year}
  {2012})}\BibitemShut {NoStop}%
\bibitem [{\citenamefont {Son}\ \emph {et~al.}(2011)\citenamefont {Son},
  \citenamefont {Amico}, \citenamefont {Fazio}, \citenamefont {Hamma},
  \citenamefont {Pascazio},\ and\ \citenamefont {Vedral}}]{son2011quantum}%
  \BibitemOpen
  \bibfield  {author} {\bibinfo {author} {\bibfnamefont {W.}~\bibnamefont
  {Son}}, \bibinfo {author} {\bibfnamefont {L.}~\bibnamefont {Amico}}, \bibinfo
  {author} {\bibfnamefont {R.}~\bibnamefont {Fazio}}, \bibinfo {author}
  {\bibfnamefont {A.}~\bibnamefont {Hamma}}, \bibinfo {author} {\bibfnamefont
  {S.}~\bibnamefont {Pascazio}}, \ and\ \bibinfo {author} {\bibfnamefont
  {V.}~\bibnamefont {Vedral}},\ }\href
  {http://stacks.iop.org/0295-5075/95/i=5/a=50001} {\bibfield  {journal}
  {\bibinfo  {journal} {EPL (Europhysics Letters)}\ }\textbf {\bibinfo {volume}
  {95}},\ \bibinfo {pages} {50001} (\bibinfo {year} {2011})}\BibitemShut
  {NoStop}%
\bibitem [{\citenamefont {Vojta}(2003)}]{ref:Vojta03-Qphase-trans}%
  \BibitemOpen
  \bibfield  {author} {\bibinfo {author} {\bibfnamefont {M.}~\bibnamefont
  {Vojta}},\ }\href {http://stacks.iop.org/0034-4885/66/i=12/a=R01} {\bibfield
  {journal} {\bibinfo  {journal} {Reports on Progress in Physics}\ }\textbf
  {\bibinfo {volume} {66}},\ \bibinfo {pages} {2069} (\bibinfo {year}
  {2003})}\BibitemShut {NoStop}%
\bibitem [{\citenamefont {Osborne}\ and\ \citenamefont
  {Verstraete}(2006)}]{ref:Osborne06-Monogamy}%
  \BibitemOpen
  \bibfield  {author} {\bibinfo {author} {\bibfnamefont {T.~J.}\ \bibnamefont
  {Osborne}}\ and\ \bibinfo {author} {\bibfnamefont {F.}~\bibnamefont
  {Verstraete}},\ }\href {\doibase 10.1103/PhysRevLett.96.220503} {\bibfield
  {journal} {\bibinfo  {journal} {Phys. Rev. Lett.}\ }\textbf {\bibinfo
  {volume} {96}},\ \bibinfo {pages} {220503} (\bibinfo {year}
  {2006})}\BibitemShut {NoStop}%
\end{thebibliography}

\providecommand{\noopsort}[1]{}\providecommand{\singleletter}[1]{#1}%

\end{document}